\documentclass[fleqn,usenatbib,usedcolumn]{mnras}
\usepackage[british]{babel}             
\usepackage{txfonts}                  
\let\la=\lesssim 
\usepackage[T1]{fontenc}                
\usepackage{graphicx}                   
\usepackage{rotating}
\usepackage{dcolumn}
\newcolumntype{d}{D{.}{.}{-1}}
\def\dhead#1{\multicolumn{1}{c}{#1}}
\def\twolines#1#2{$\kern-6pt\Big\{ {\textrm{#1\hfill}\atop\textrm{#2\hfill}}$}
\def\vpad{{\Large$\mathstrut$}}
\usepackage{multirow}
\graphicspath{{.}{./FIGS/}}

\hypersetup{pdfauthor={I. H. Whittam, J. M. Riley, D. A. Green, M. L. Davies, T. M. O. Franzen, C. Rumsey, M. P. Schammel and E. M. Waldram},
               pdftitle={10C continued: a deeper radio survey at 15.7 GHz},
               pdfkeywords={galaxies: active, radio continuum: galaxies, catalogues, surveys},
               bookmarksnumbered=true}
\hypersetup{colorlinks=true,
            linkcolor=blue,
            citecolor=blue,
            filecolor=blue,
            urlcolor=blue}

\volume{{\rm in press}}

\title[10C continued: a deeper radio survey at 15.7 GHz]{10C continued: a deeper radio survey at 15.7 GHz}

\author[I.~H.~Whittam et al.]{\parbox{\textwidth}{I.~H.~Whittam$^{1,2}$\thanks{email:
\texttt{i.whittam@mrao.cam.ac.uk}}, J.~M.~Riley$^2$, D.~A.~Green$^2$, M.~L.~Davies$^2$, T.~M.~O.~Franzen$^3$, C.~Rumsey$^2$, M.~P.~Schammel and E.~M.~Waldram$^2$.}\vspace{0.4cm}\\
   $^{1}$Physics Department, University of the Western Cape, Bellville 7535, South Africa\\
   $^{2}$Astrophysics Group, Cavendish Laboratory, 19 J.~J.~Thomson Avenue, Cambridge CB3 0HE\\
   $^{3}$ International Centre for Radio Astronomy Research (ICRAR), Curtin University, Bentley, WA 6102, Australia}

\date{Accepted ---; received ---; in original form ---}

\pagerange{\pageref{firstpage}--\pageref{lastpage}}

\pubyear{2015}
\begin{document}

\label{firstpage}

\maketitle

\begin{abstract}
We present deep 15.7-GHz observations made with the Arcminute Microkelvin Imager Large Array in two fields previously observed as part of the Tenth Cambridge (10C) survey. These observations allow the source counts to be calculated down to 0.1~mJy, a factor of five deeper than achieved by the 10C survey. The new source counts are consistent with the extrapolated fit to the 10C source count, and display no evidence for either steepening or flattening of the counts. There is thus no evidence for the emergence of a significant new population of sources (e.g.\ starforming) at 15.7~GHz flux densities above 0.1~mJy, the flux density level at which we expect starforming galaxies to begin to contribute. Comparisons with the \citeauthor{2005A&A...431..893D} model and the SKADS Simulated Sky show that they both underestimate the observed number of sources by a factor of two at this flux density level. We suggest that this is due to the flat-spectrum cores of radio galaxies contributing more significantly to the counts than predicted by the models.
\end{abstract}


\section{Introduction}\label{section:intro}

The high-frequency radio sky ($\nu \gtrsim 10$~GHz) has been much less widely studied than the population at lower radio frequencies (e.g.\ 1.4~GHz), mainly due to the increased time required to survey an area to an equivalent depth at higher frequencies. In recent years several high-frequency surveys have been conducted, albeit with higher flux density limits than surveys at lower frequencies. \citet{2003MNRAS.342..915W,2010MNRAS.404.1005W} carried out the Ninth Cambridge (9C) survey at 15~GHz using the Ryle Telescope. This survey covers 520~deg$^2$ to a completeness limit of $\approx$ 25~mJy, and was the first high-frequency survey to cover a significant proportion of the sky. A series of deeper regions were also observed \citep{2010MNRAS.404.1005W}, with 115~deg$^2$ complete to $\approx$ 10~mJy and 29~deg$^2$ complete to $\approx$ 5.5~mJy. 

The whole Southern sky has been surveyed by the Australia Telescope 20 GHz (AT20G) survey \citep{2011MNRAS.412..318M}, which has a flux density limit of 40 mJy and is 93 per cent complete above 100 mJy. This survey is complementary to the 9C survey as it covers a larger area at a shallower flux density. More recently, the AT20G-deep pilot survey \citep{2014MNRAS.439.1212F} surveyed 5~deg$^2$ to a completeness level of 2.5~mJy.

The Tenth Cambridge (10C) survey was observed with the Cambridge Arcminute Microkelvin Imager (AMI; \citealt{2008MNRAS.391.1545Z}) at 15.7~GHz. The observations, mapping and source extraction are described in \citet{2011MNRAS.415.2699F} (hereafter Paper F) and the source counts and spectral properties are presented in \citet{2011MNRAS.415.2708D} (hereafter Paper D). The 10C survey is complete to 1~mJy in ten different fields covering a total of $\approx 27 \textrm{ deg}^2$; deep areas covering $\approx 12 \textrm{ deg}^2$, contained within these fields, are complete to 0.5~mJy, making the 10C survey the deepest high-frequency radio survey published to date. 

\citet{2013MNRAS.429.2080W} used data at a range of frequencies to study the spectral indices of 10C sources in the Lockman Hole. They found a significant change in spectral index with flux density; the median spectral index $\alpha$ (where $S \propto \nu^{-\alpha}$ for a source with flux density $S$ at frequency $\nu$) calculated between 1.4 and 15.7~GHz changes from 0.75 for sources with $S_{15.7~\rm GHz} > 1.5$~mJy to 0.08 for sources with $S_{15.7~\rm GHz} < 0.8$~mJy. This shows that there is a population of flat-spectrum sources emerging below 1~mJy; \citeauthor{2013MNRAS.429.2080W} suggest that this may be due to the cores of Fanaroff and Riley type I (FRI; \citealt{1974MNRAS.167P..31F}) sources becoming dominant at 15.7~GHz.

There have been several attempts to model the high-frequency radio sky, often extrapolating from lower frequencies. Early evolutionary models of radio sources \citep{1990MNRAS.247...19D,1999MNRAS.304..160J,1998MNRAS.297..117T} successfully fitted the available data at frequencies $\lesssim$ 10 GHz down to flux densities of a few mJy. More recently, \citet{2005A&A...431..893D} produced a model of the radio source counts at frequencies $\gtrsim$ 5 GHz (up to 30 GHz) which successfully fitted the data available at the time. The \citeauthor{2005A&A...431..893D} model splits the sources into flat and steep-spectrum populations, with the flat-spectrum population further divided into flat-spectrum radio quasars and BL Lacs, and determines the epoch-dependent luminosity function for each population. Starforming galaxies, GHz-peaked spectrum (GPS) sources and objects in the late stages of AGN evolution are also included in the model.

A more recent model by \citet{2011A&A...533A..57T} used physically grounded models to extrapolate the 5-GHz source count, which is well known observationally, to higher frequencies. They focus on the spectral behaviour of blazars and compare three different models which treat flat-spectrum sources differently. This scheme is successful at high flux densities but does not accurately reproduce the observed 15 GHz source count below $\approx$ 10 mJy. The number of sources is significantly underestimated (by a factor of $\sim 2$), indicating that the properties of these sources are not well understood, largely due to the complexity and diversity of the high-frequency spectra of individual sources. 

\cite{2008MNRAS.388.1335W,2010MNRAS.405..447W} have produced a semi-empirical simulation of the extragalactic radio continuum sky (the SKADS Simulated Sky; S$^3$). The simulation splits the radio sources into separate populations: FRII sources, FRI sources, radio-quiet AGNs and starforming galaxies, which are split into quiescent starforming and starbursting galaxies. The observed (and extrapolated) radio continuum luminosity functions are used to generate a catalogue of $\approx$ 320 million simulated sources. This simulation covers 20 $\times$ 20 deg$^2$ out to a cosmological redshift of $z=20$ and down to a flux density of 10 nJy at 151, 610 MHz, 1.4, 4.86 and 18 GHz. \citet{2013MNRAS.429.2080W} showed that the simulation fails to reproduce the spectral index distribution of 10C sources, dramatically under-predicting the number of flat spectrum sources.

Thus, although good progress has been made in recent years in modelling aspects of the extragalactic radio source population, the high-frequency radio source population at low flux densities is poorly described. To understand the nature of the faint, high-frequency population and constrain the models better observations of the faint, high-frequency sky are required. In this paper we present new, deep observations at 15.7~GHz in two 10C fields, which when combined with existing 10C data enable the source counts to be constrained down to 0.1~mJy, a factor of five deeper than the 10C source count.

This paper is laid out as follows. Section~\ref{section:obs} describes the observations and data reduction, and Section~\ref{section:source_catalogue} details the methods used to produce the source catalogue. Section~\ref{section:checks} discusses the effects of flux density variability. The catalogue is used to derive the source counts in Section~\ref{section:source_counts} and these results are discussed in Section~\ref{section:sc_discussion} before some brief conclusions are presented in Section~\ref{section:conclusions}.

\section{Observations and data reduction}\label{section:obs}

The 15.7-GHz observations were conducted between August 2008 and July 2014 using the Arcminute Microkelvin Imager Large Array (AMI LA), located near Cambridge, UK. The AMI LA consists of eight 13-m antennas with baselines of 18 to 110~m, giving it a primary beam of 5.5~arcmin and a resolution of 30~arcsec.

For the 10C survey the LA observations were made of 10 fields of different sizes, covering a total of 27~deg$^2$. Each field consists of an outer region complete to 1~mJy beam$^{-1}$, and an inner deeper region complete to 0.5~mJy beam$^{-1}$.

\begin{figure}
\centerline{\includegraphics[width=\columnwidth]{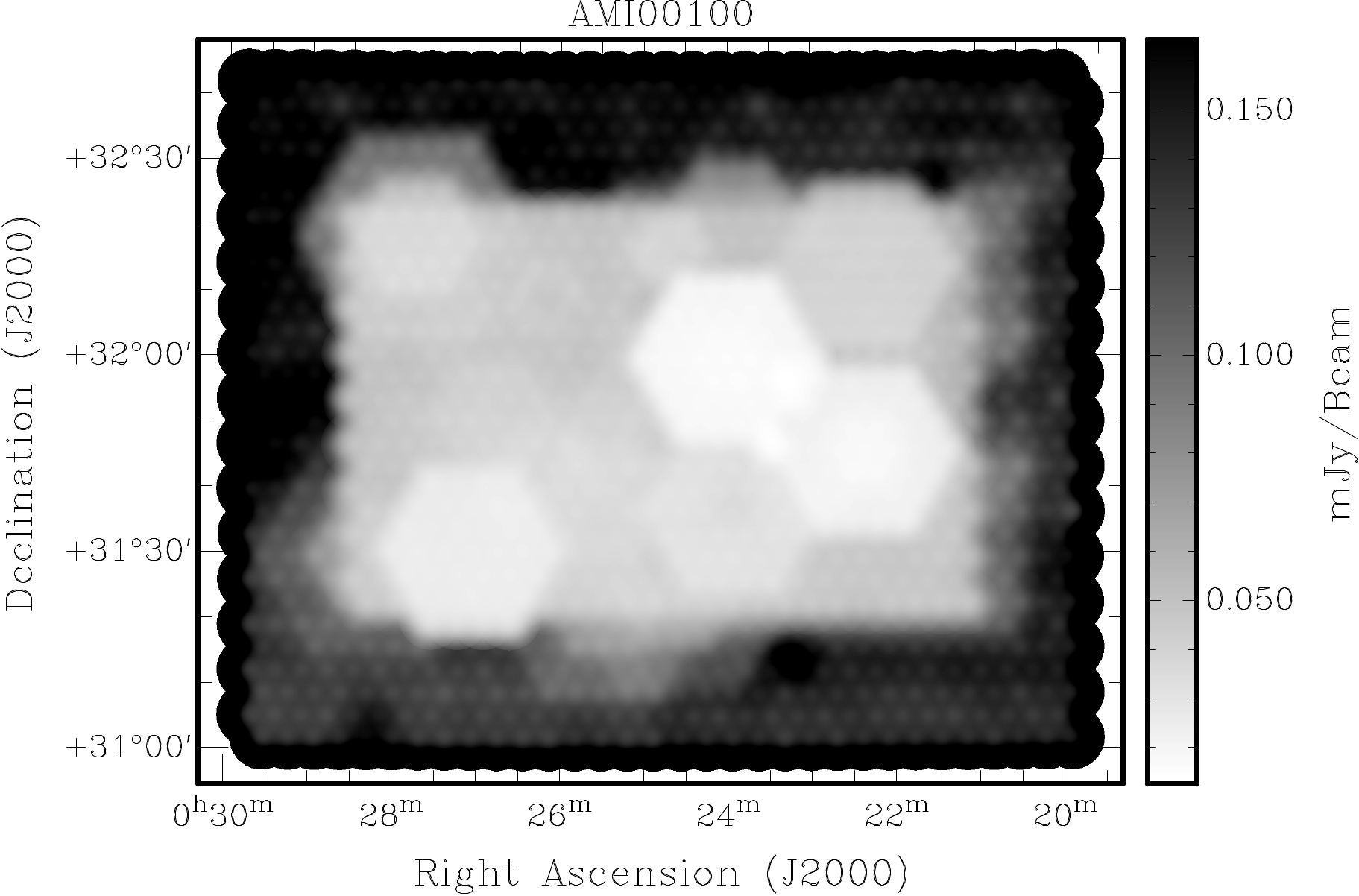}}
\caption{The noise in the 15.7-GHz AMI map of the AMI001 (J0024+3152) field. }\label{fig:AMI1_noise}
\end{figure}

\begin{figure}
\centerline{\includegraphics[angle=270,width=\columnwidth]{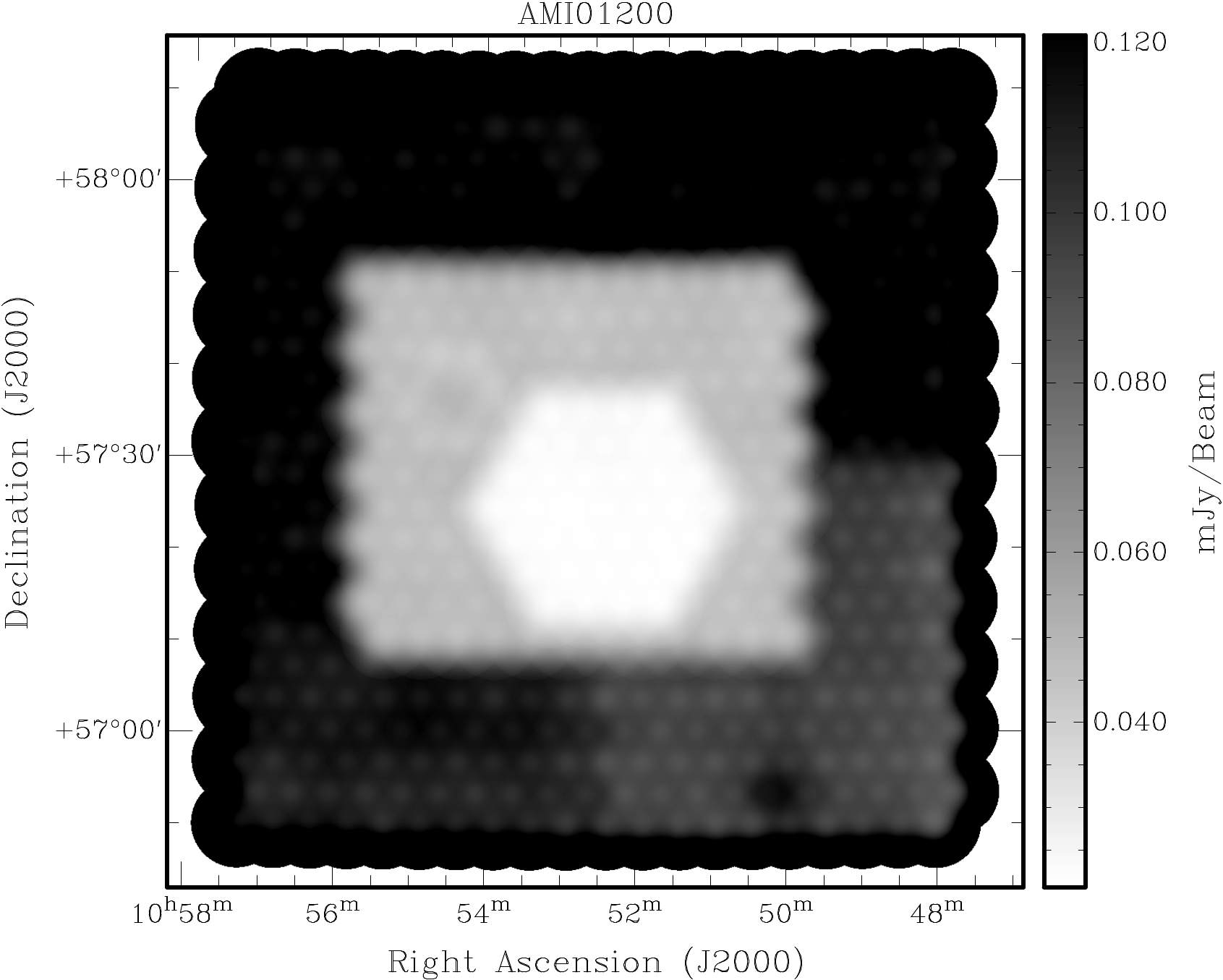}}
\caption{The noise in the 15.7-GHz AMI map of the Lockman Hole (J1052+5730) field. }\label{fig:LH_noise}
\end{figure}

Here we describe further observations of two of the 10C fields: the AMI001 (J0024+3152) field and the Lockman Hole field (J1052+5730), details of which are given in Table \ref{tab:regions}. From these two fields we selected sub-fields that we observed further, using a rastering technique in the same way as the 10C survey observations, with hexagonal rasters of 37 pointings each separated by 4 arcmin. In the AMI001 field, 450 hours of new observations have been made in several hexagons (each consisting of 37 pointings), and the rms noise in the centre of the two deepest hexagons is $\sim 16~\muup \rm Jy~beam^{-1}$. The noise map for the AMI001 field is shown in Fig. \ref{fig:AMI1_noise}. In the Lockman Hole field, 300 hours of additional observations were carried out in one 37-pointing hexagon. The rms noise in this hexagon is $\sim 21~\muup$Jy beam$^{-1}$, as shown by the noise map in Fig. \ref{fig:LH_noise}. 

\begin{table*}
\caption{Summary of the different regions in the two fields.}\label{tab:regions}
\centering
\begin{tabular}{lllld}\\\hline
Region & Comments & Area 	  & Observing time & \dhead{approx. rms} \\
	   &  		  & / deg$^2$ & / hours        & \dhead{/ $\muup$Jy} \\\hline
\multicolumn{5}{l}{AMI001 (J0024+3152)}\\\hline
10C shallow & Complete to 1 mJy   & 3.56 & \multirow{2}{*}{1000} & 150 \\
10C deep    & Complete to 0.5 mJy & 1.69 &  & 45\\
10C ultra-deep    & This work	-- deepest obs. in two hexagons centred on 	& 0.38	& additional 450 & 16\\
			& $00^{\rm h} 22^{\rm m} 19^{\rm s}, +31^{\circ} 46' 09.7''$ and $00^{\rm h} 24^{\rm m} 00^{\rm s}, +31^{\circ} 59' 29.5''$ & & & \\\hline
\multicolumn{5}{l}{Lockman Hole (J1052+5730)}\\\hline
10C shallow & Complete to 1 mJy   & 1.78 & \multirow{2}{*}{450} & 120 \\
10C deep    & Complete to 0.5 mJy & 0.54 &  & 45\\
10C ultra-deep    & This work	-- hexagon centred on $0^{\rm h} 52^{\rm m} 22^{\rm s}, +57^{\circ} 24' 55.0''$	 & 0.18 & additional 300 & 21\\\hline
\end{tabular}
\end{table*}

All raw data files were reduced with the up-to-date reduction pipeline in the AMI in-house software package \textsc{reduce} (Paper F). This new pipeline was developed from that used for the 10C survey with enhanced methods for flagging interference and produces consistent results. $uv$-fits files are written out from \textsc{reduce} and concatenated to produce a single $uv$-fits file for each field. Mapping is carried out in \textsc{aips} using the same imaging pipeline as that for the 10C survey, with automated \textsc{clean} procedures and the \textsc{imean} task used to estimate the noise. The pixel size in the image is $5 \times 5$~arcsec. Details of the data reduction and imaging procedure can be found in Paper F.

\section{The source catalogue}\label{section:source_catalogue}

\subsection{Source fitting}\label{section:source_fitting}

The source fitting was performed using the Cambridge in-house software \textsc{source\_find} in the same way as for the 10C survey in order to make the new catalogue as comparable as possible to the 10C catalogue. The procedure implemented in \textsc{source\_find} is described briefly here, full details being given in Paper F. The images were used in conjunction with the noise maps shown in Figs. \ref{fig:AMI1_noise} and \ref{fig:LH_noise} to identify component `peaks' above a given signal-to-noise ratio $\gamma$ (in this case $\gamma =  4.62$\footnote{This value is used, instead of 5$\sigma$, to take account of the 1.082 phase error correction factor applied to the flux densities (see Section~\ref{section:flux_values} for details).}). Initially, all pixels greater than $0.6 \gamma \sigma$ (where $\sigma$ is the local noise) were identified to ensure all peaks greater than $\gamma \sigma$ after interpolation are found. The position and flux density of each peak (RA$_{\rm pk}$, Dec$_{\rm pk}$ and $S_{\rm pk}$) was then found by interpolation between the pixels, and any peaks less than $\gamma \sigma$ were discarded. The error on the peak flux density ($\Delta S_{\rm pk}$) is taken to be the thermal noise error $\sigma_{\rm n}$ combined in quadrature with the five per cent calibration error, $\Delta S_{\rm pk} = \sqrt{\sigma_{\rm n}^2 + (0.05S_{\rm pk})^2}$. The integration area, consisting of contiguous pixels down to a lowest contour value of 2.5$\sigma$, was then calculated for each component. If more than one peak lay inside the same integration area the sources were classified as a `group'.

The \textsc{aips} task \textsc{jmfit} was then used to fit a 2D elliptical Gaussian to each component. This was used to estimate the integrated flux density, position and angular size (RA$_{\rm in}$, Dec$_{\rm in}$, $S_{\rm in}$ and $e_{\rm maj}$) for each component. The error on the integrated flux density ($\Delta S_{\rm in}$) was estimated as the error due to thermal noise ($\sigma_{\rm n}$) combined in quadrature with the five per cent calibration error, $\Delta S_{\rm in} = \sqrt{\sigma_{\rm n}^2 + (0.05S_{\rm in})^2}$.

\subsection{Exclusion zones}\label{section:exclusion-zones}

The maps display an increase in noise close to bright ($>15$~mJy) sources due to amplitude, phase and deconvolution errors. Because this noise is non-gaussian it is often not included in the noise maps, leading to spurious detections close to bright sources. For this reason, an `exclusion zone' was defined around any source with a peak flux density greater than 15~mJy during the source fitting process, and any source detected within this zone was excluded from the catalogue. As derived empirically in Paper D, each exclusion zone is a circle centred on the bright source, with a radius defined by:
\begin{equation}
r_{\rm e} = 12 \left({\frac{S_{\rm pk}}{250~\rm{mJy}}}\right)^{1/2} \rm{arcmin},
\end{equation}
where $S_{\rm pk}$ is the peak flux density of the bright source. The positions and radii of these exclusion zones are shown in Table \ref{tab:zones}.

\begin{table}
\caption{The positions of the centres of the exclusion zones around bright sources and their radii.}\label{tab:zones}
\centering
\begin{tabular}{ccc}\\\hline
RA & Dec & Radius / arcmin\\\hline
00:29:33.7 &   +32:44:52   &     4.19\\
00:23:09.8 &   +31:14:00   &     3.80\\
00:21:29.8 &   +32:26:58   &     3.41\\
00:20:50.4 &   +31:52:28   &     3.30\\
00:28:10.5 &   +31:03:46   &     3.20\\
00:29:20.4 &   +32:16:54   &     3.06\\
10:50:07.1 &   +56:53:37   &     3.35\\
10:52:25.1 &   +57:55:07   &     3.24\\
10:54:26.9 &   +57:36:48   &     3.69\\\hline
\end{tabular}
\end{table}

\subsection{Final flux density values}\label{section:flux_values}

The deep AMI images do not have a signal-to-noise high enough to be self-calibrated, so a correction factor was applied to all flux density values to account for the residual phase errors before inclusion into the final catalogue (Paper D). The correction factor used here is the same as the one used for the 10C survey, which was estimated by comparing the peak flux densities of bright, unresolved sources in the 10C raster maps
with their peak flux densities in self-calibrated, pointed maps. All peak flux densities were therefore multiplied by 1.082. The use of this correction factor is the reason for carrying out the source-fitting down to 4.62$\sigma$ -- a source near the survey limit which should be detected at 5$\sigma$ will only be detected at $5\sigma / 1.082 = 4.62\sigma$. In total, 358 sources were detected in the AMI001 field and 134 in the Lockman Hole field (including the original shallower areas from Paper D as well as the new deep areas).

A source was considered to be extended if the major axis of the deconvolved Gaussian ($e_{\rm maj}$) is larger than a critical value $e_{\rm crit}$ (see Paper F), where
\begin{equation}
e_{\rm crit} =  \left\{ 
						\begin{array}{l l}
						3.0\, b_{\rm maj}\, \rho^{-1/2}  & \textrm{if} ~3.0\, b_{\rm maj} \, \rho^{-1/2} > 25.0~\textrm{arcsec},\\
						25.0~\rm arcsec				& \textrm{otherwise},\\
						\end{array}\right.
\label{eqn:ecrit}
\end{equation}
where $b_{\rm maj}$ is the major axis of the restoring beam and $\rho = S_{\rm pk} / \sigma_{\rm n}$ (i.e.\ the signal-to-noise ratio). Sources with $e_{\rm maj} > e_{\rm crit}$ were classified as extended (flag E), otherwise the source was considered point-like (flag P). In total, there are 24 sources classified as extended in the AMI001 field and six in the Lockman Hole field. 

For sources classified as extended, the integrated flux density is used, and for those classified as point-like, the peak flux density is used. These values are listed as the `best flux' in the catalogue, and are the values used for determining the source count in Section~\ref{section:source_counts}.

\subsection{Completeness}

\begin{figure}
\centerline{\includegraphics[width=\columnwidth]{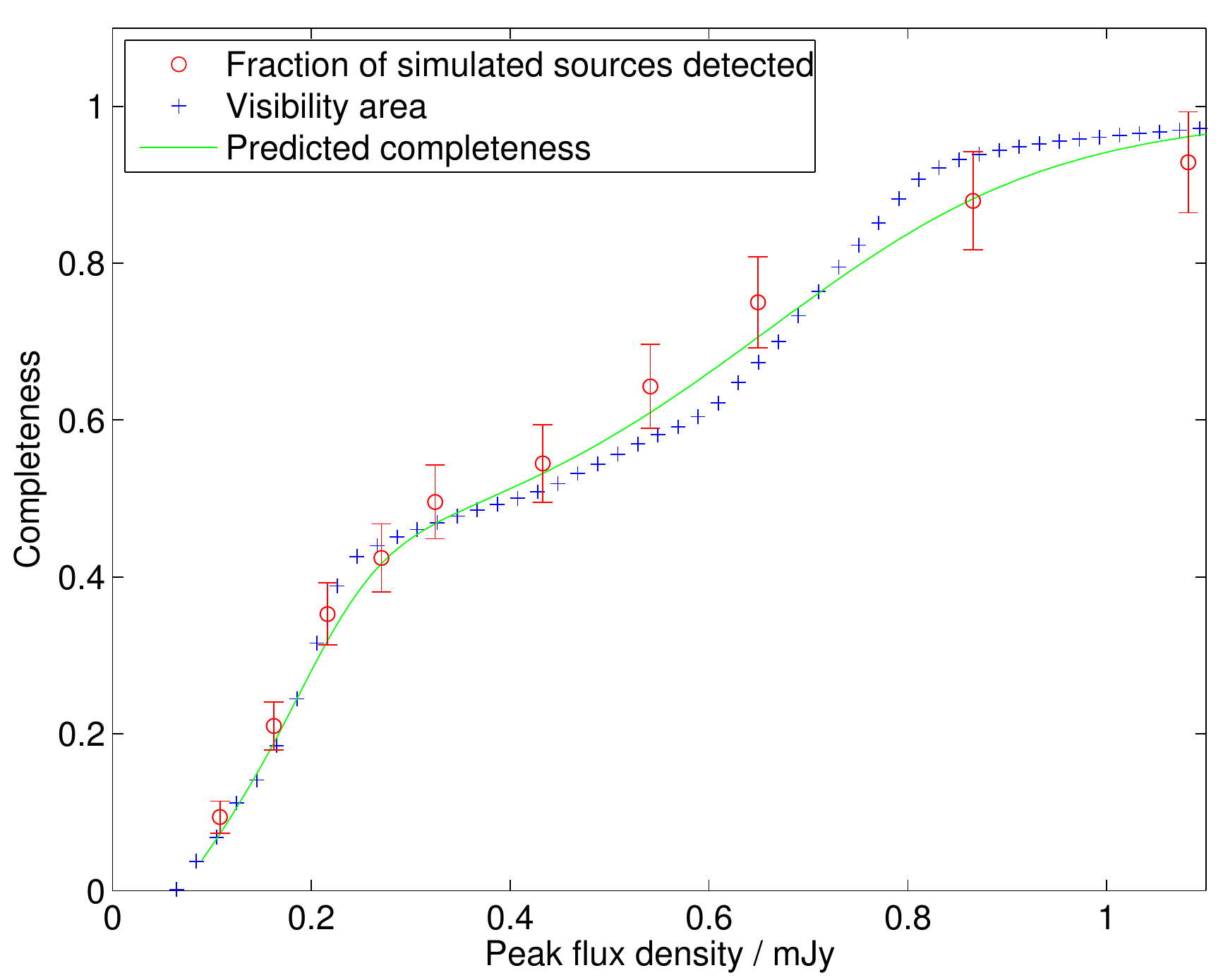}}
\caption{Different methods used to estimate completeness in the AMI001 field. The fractions of simulated sources detected at a given flux density are shown by red circles (error bars represent Poisson errors). The completeness curve estimated from the noise map using equation \ref{eqn:completeness} is shown by the green line. The blue crosses show the visibility area used to calculate the source counts in Section~\ref{section:source_counts}.}\label{fig:completeness}
\end{figure}

The completeness of the survey was estimated by inserting 250 sources of equal flux density into the AMI001 map in random positions. The simulated sources were ideal point sources with flux density $S$ and were added to the map in the image plane using the \textsc{aips} task \textsc{immod}; no sources were placed in the border 50 pixels wide at the edge of the map, as many of these pixels are blank (the blank pixels were excluded in area and source count calculations). Sources which lay within 2 arcmin of another simulated source were removed to avoid the simulated sources contaminating the flux measurement of the adjacent source, leaving 224 simulated sources. The source fitting was then performed in exactly the same way as described in Section~\ref{section:source_fitting}. A simulated source was considered to be detected if there was a source in the output catalogue within 30 arcsec of the simulated source position. This was repeated several times using a range of flux densities $0.1 < S/\rm{mJy} < 1$ (keeping the same positions each time) to estimate the completeness as a function of flux density. The fraction of simulated sources detected as a function of flux density is shown in Fig. \ref{fig:completeness}. The flux densities of the simulated sources are multiplied by 1.082 (the correction factor applied to the final catalogue to account for phase errors, see Section~\ref{section:flux_values}) before inclusion on this plot.

The completeness can also be estimated from the noise map assuming a Gaussian noise distribution. The probability of detecting a source of peak flux density $S$ above 5$\sigma$ is given by
\begin{equation}
P(S) = \int_{5 \sigma}^\infty \frac{1}{\sqrt{2 \pi \sigma^2}} \exp \left( - \frac{(X - S)^2}{2 \sigma^2}\right) {\rm d}X.
\end{equation}
In reality the noise varies across the map; this can be taken into account by averaging the probabilities of detecting a source at each pixel position in the noise map. As the source fitting was in fact carried out to 4.62$\sigma$ and the flux densities were multiplied by 1.082 after source extraction, this equation becomes
\begin{equation}
P_i(S) = \int^\infty_{4.62\sigma_i} \frac{1}{\sqrt{2 \pi \sigma_i^2}} \exp \left(- \frac{\left( X - \frac{S}{1.082}\right)^2}{2 \sigma_i^2}\right) {\rm d}X,
\label{eqn:completeness}
\end{equation}
where $\sigma_i$ is the value of the noise map at the $i^{\rm th}$ pixel. The outer 50 pixels of the noise map are excluded from this calculation so that the results are directly comparable to the completeness estimated using simulated sources.

The results of both methods used to estimate completeness are shown in Fig. \ref{fig:completeness}. A similar analysis was performed in Paper D to estimate the completeness of the 10C survey. Paper D found that at the lowest flux densities probed by their survey ($S \sim 0.5~\rm mJy$) the fraction of detected sources from the simulation was slightly higher than predicted by the completeness curve, while at higher flux densities the fraction of detected sources was slightly lower than predicted. The same effect is visible here, with the fraction of detected sources slightly higher than predicted for $S < 0.8$~mJy and lower than predicted for $S > 0.8$~mJy. Paper D suggests several factors which may contribute to this effect. One is source confusion, which will increase the completeness at low flux densities, as two sources below the completeness limit may lie sufficiently close together to be detected as one source; in contrast, at higher flux densities confusion will reduce the completeness as a source may not be detected if it lies too close to a brighter source. Confusion therefore prevents the completeness from reaching 100 per cent as quickly as predicted. 

The `visibility area' \citep{1973A&A....23..171K} is also plotted in Fig. \ref{fig:completeness}. This is the fraction of the total area over which a source of flux density $S_i$ should be detectable, i.e. the fraction of the area with 5$\sigma_{\rm local} < S_i$. This is used to calculate the source count and is discussed in more detail in Section~\ref{section:source_counts}.

\subsection{Reliability}

Assuming that the noise in the map is Gaussian, the expected number of false positives simply due to the noise can be calculated (the noise is unlikely to be Gaussian close to bright sources, but as exclusion zones are placed around sources with $S>15$~mJy (see Section~\ref{section:exclusion-zones}) this assumption is valid). The total area of the two fields containing the deeper observations is 5.3~deg$^2$ and the synthesised beam area is $\approx 700$~arcsec$^2$, so the number of beams covering the two fields is $\approx 100,000$. The probability that a value drawn from a Gaussian distribution is more than 4.62 standard deviations away from the mean is $\approx 1.9 \times 10^{-6}$, so the expected number of false positives is $\approx 0.2$. We therefore do not expect reliability to be an issue for this survey.

\subsection{Multiple sources}

There are 38 sources in the AMI001 field which are classified as being part of a `group', indicating that their fitted Gaussians overlap. Twelve of these sources form four triples, the remaining 26 sources form 13 pairs, meaning that there are 17 groups in total. The separation between the sources in these 17 groups ranges from 27 to 126 arcsec, with an average separation of 49 arcsec. The source counts can be used to estimate the number of sources that would be expected to be overlapping due to confusion alone. Integrating the 10C source counts between $0.1 < S/{\rm mJy} < 25$ (which involves extrapolation from 0.5 to 0.1~mJy), $\approx 7.2 \times 10^5$ sources per steradian are expected in this flux density range. There are a total of 358 sources in the AMI001 field, so the total area within 49 arcsec of a source is $358 \times A_{\rm circ} \approx 6.4 \times 10^{-5}$ steradians (where $A_{\rm circ}$ is the area of a circle with a radius of 49 arcsec). Thus $\approx 46$ sources are expected to lie within 49 arcsec of another source and would consequently be classified as overlapping. This is higher than the number of overlapping sources observed, which is probably due to the fact that sources as faint as 0.1~mJy (the lower limit on flux density used in this calculation) can only be detected in part of the image. This number can therefore be viewed as an upper limit on the expected number of overlapping sources. Repeating this calculation with a detection limit of 0.5~mJy shows that we would expect $\approx 10$ sources to be overlapping; we therefore expect between 10 and 46 overlapping sources in the field due to confusion alone. This estimate assumes no source clustering, which would increase the number of overlapping sources. 

This analysis suggests that many of the overlapping sources detected in these observations are due to confusion, rather than being genuine multiple sources. No attempt is made therefore to combine the flux densities of the overlapping sources into a single source; they are listed as separate entries in the source catalogue, but flagged as being part of a group.

\subsection{The final catalogue}\label{section:catalogue}

\begin{figure*}
\centerline{\includegraphics[width=\columnwidth]{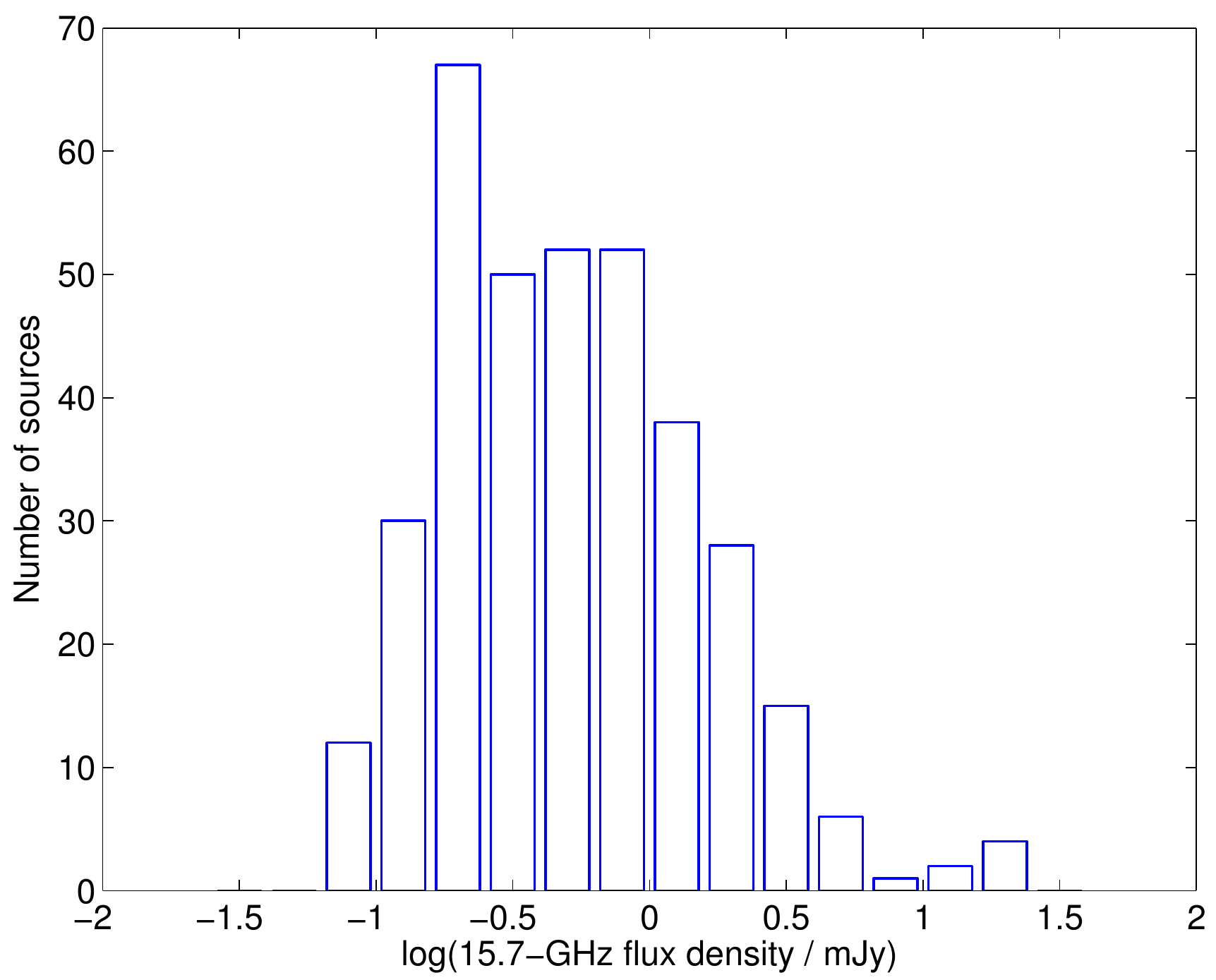}
            \quad
            \includegraphics[width=\columnwidth]{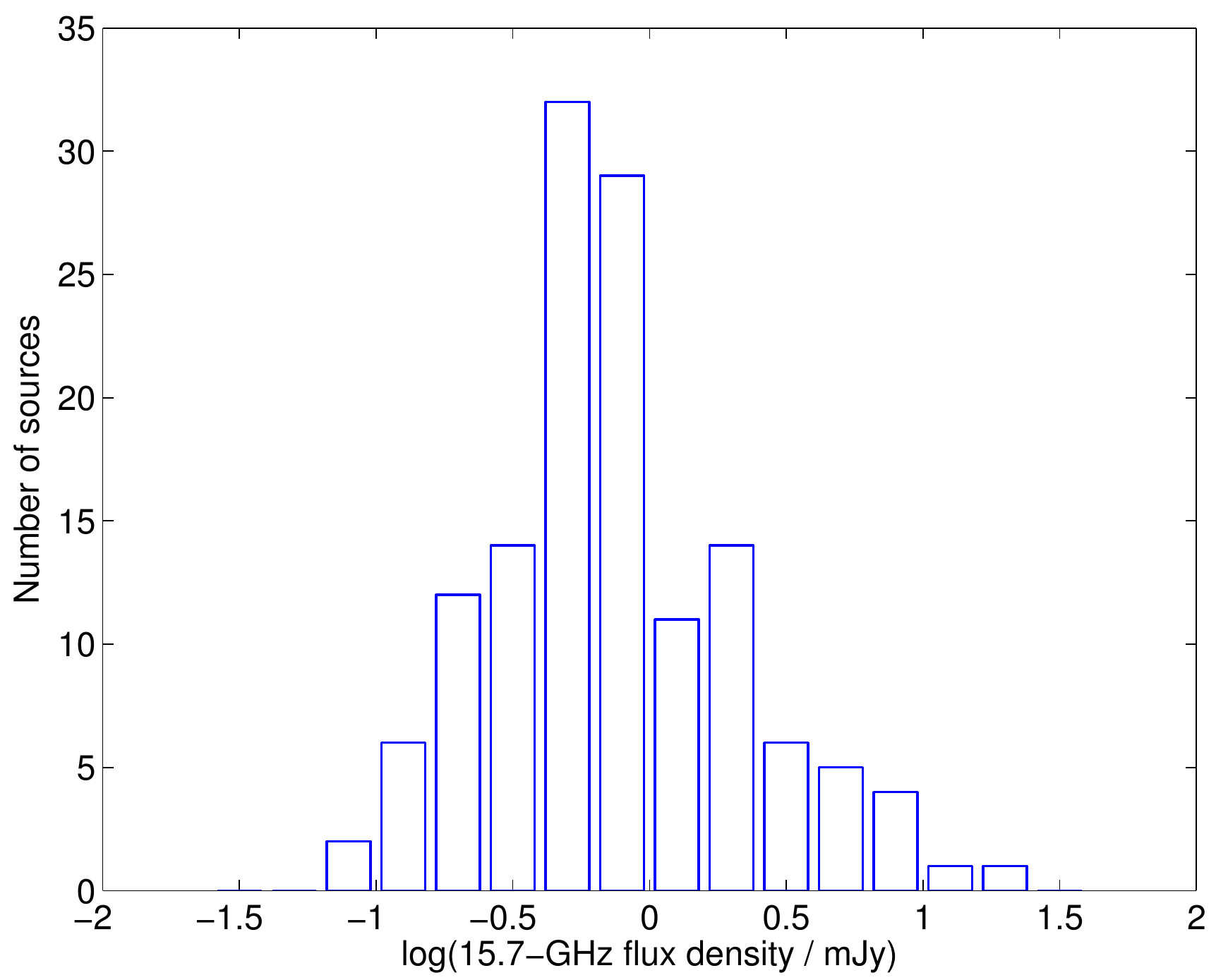}}
\caption{Flux density distributions of sources in the AMI001 field catalogue (left) and the Lockman Hole field catalogue (right).}\label{fig:flux_hist}
\end{figure*}

The final source catalogue contains 358 sources in the AMI001 field with flux densities $0.088 < S/{\rm mJy} < 34$ and 134 sources in the Lockman Hole field with flux densities $0.15 < S/{\rm mJy} < 22$. The flux density distributions of the sources in these two catalogues are shown in Fig. \ref{fig:flux_hist}. Twenty-four of the sources in the AMI001 field (7 per cent) and six in the Lockman Hole field (4 per cent) are classified as extended. The source catalogues are available online, the parameters which appear in these catalogues are described in Table \ref{tab:catalogue}.

\begin{table}
\caption{The parameters which appear in the source catalogues available online. Full details of all parameters are given in Paper F.}\label{tab:catalogue}
\centering
\begin{tabular}{ll}\\\hline
Source & 10C source designation J2000 \\
Group  & 10C group designation J2000 \\
$\alpha_{\rm pk}$ & RA (peak), in h, m, s (J2000) \\
$\delta_{\rm pk}$ & Dec. (peak), in $^\circ,',''$ (J2000)\\ 
$S_{\rm pk}$ & Peak flux density, in mJy beam$^{-1}$\\
$\delta S_{\rm pk}$ & Error on the peak flux density, in mJy beam$^{-1}$\\
$\alpha_{\rm in}$ & RA (fitted peak), in h, m, s \\
$\delta_{\rm in}$ & Dec. (fitted peak), in $^\circ,',''$\\	 
$S_{\rm in}$ & Integrated flux density, in mJy \\
$\delta S_{\rm in}$ & Error on the integrated flux density, in mJy \\
$e_{\rm crit}$ & Critical component size, in arcsec \\
$e_{\rm maj}$ & Major-axis after deconvolution, in arcsec \\
$e_{\rm min}$ & Minor-axis after deconvolution, in arcsec \\
$e_{\theta}$ & Position angle after deconvolution, in $^\circ$,\\ 
             & measured from north to east \\
$t$ & Source type (P = point-like, E = extended) \\
Flag  & A star indicates that the approximation error for\\
      & the point-source response is significant \\\hline
\end{tabular}
\end{table}

\section{The effects of variability}\label{section:checks}

As these observations were made over a six year period, they can be used to investigate the effects of variability on the flux densities. This also provides a useful check on the map-making process and the catalogue itself.

\subsection{Comparison with the original 10C catalogue}

The source catalogue produced  in the AMI001 field was compared with the original 10C catalogue. There are 358 sources in the new catalogue and 290 in the original 10C catalogue; there are 88 sources found in the new catalogue which are not found in the 10C catalogue and 20 sources in the 10C catalogue which are not in the new catalogue. In each 10C field two complete areas were defined; one area complete to 1~mJy which encompasses most of the field (the `shallow' area), and one area complete to 0.5~mJy (the `deep' area) which is contained within the shallow area. The 20 sources not found in the new map are all at the edge of the map, outside the complete area.
 
All of the sources in the new, deep areas which are not present in the 10C catalogue have flux densities less than 0.5~mJy (the completeness limit of the 10C survey), as expected. There are three sources in the field which are included in the deep complete 10C catalogue, and therefore have flux densities greater than 0.5~mJy, which have flux densities less than 0.5~mJy in the new catalogue. However, due to the relatively large errors at low signal-to-noise levels, these new flux density values are all consistent with 0.5~mJy within the errors.

\begin{figure*}
\centerline{\includegraphics[width=\columnwidth]{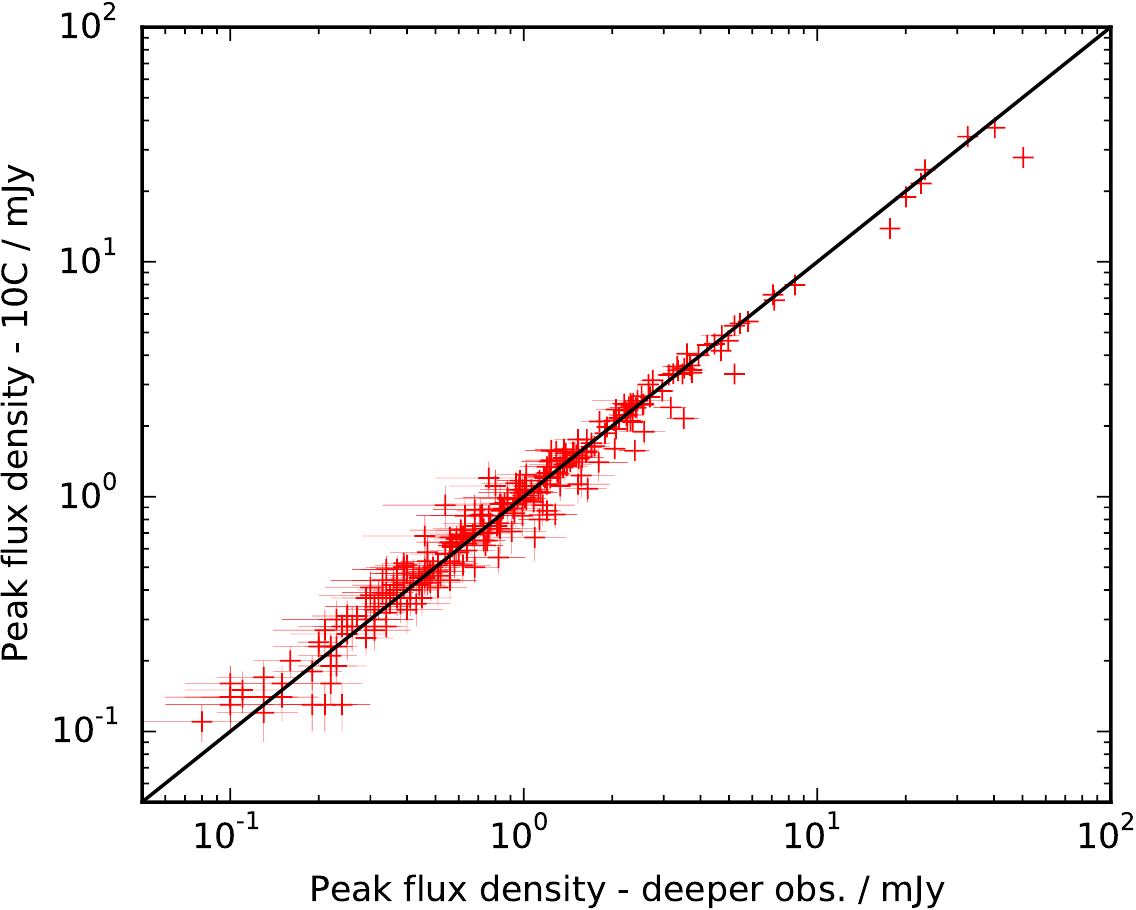}
            \quad
            \includegraphics[width=\columnwidth]{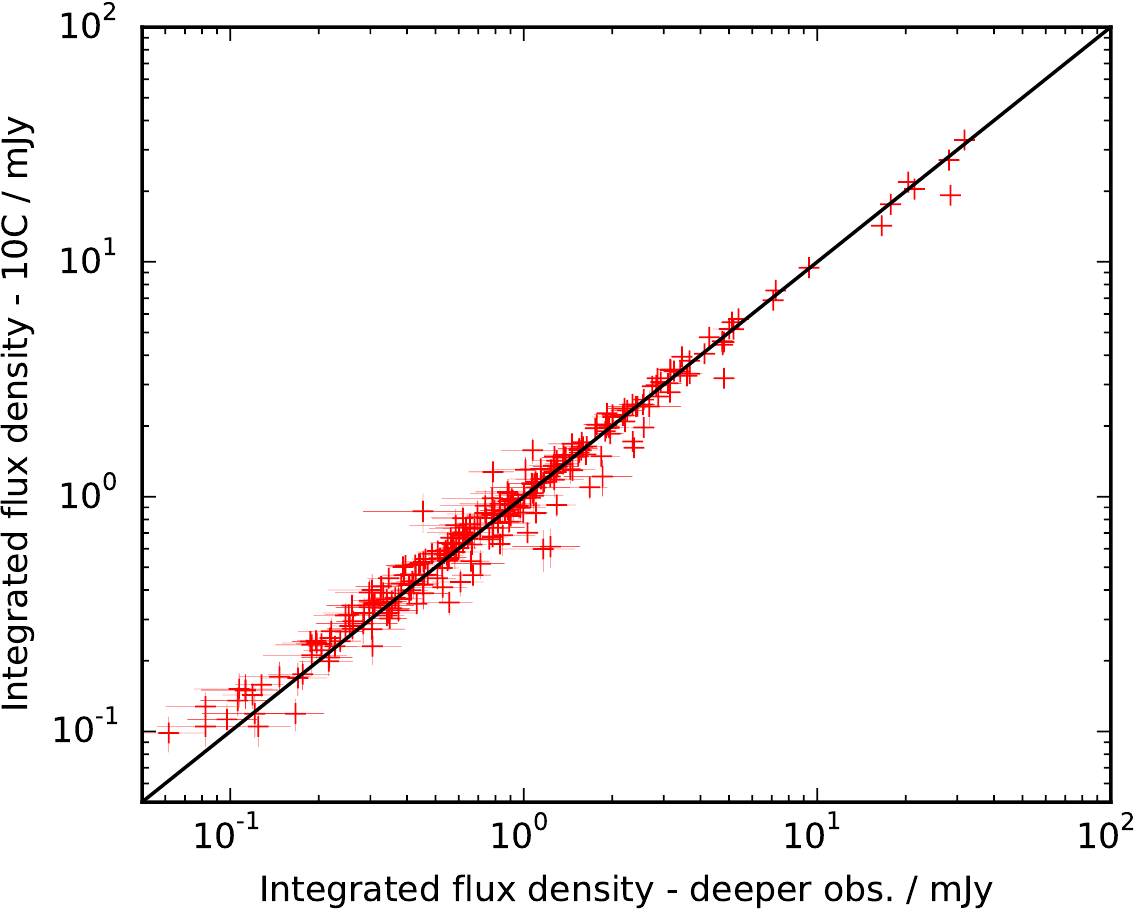}}
\caption{Comparison of peak (left panel) and integrated (right panel) flux densities in the 10C and deeper catalogues in the AMI001 field. The black line indicates where the two flux densities are equal.}\label{fig:10C_new_comparison}
\end{figure*}

Fig. \ref{fig:10C_new_comparison} shows a comparison of the peak and integrated flux densities of sources which appear in both the original 10C catalogue and the new, deeper catalogue in the AMI001 field. The flux densities from the two catalogues are generally in good agreement, as expected given that two-thirds of the data came from 10C.

\subsection{Splitting the data in half}

\begin{figure}
\centerline{\includegraphics[width=\columnwidth]{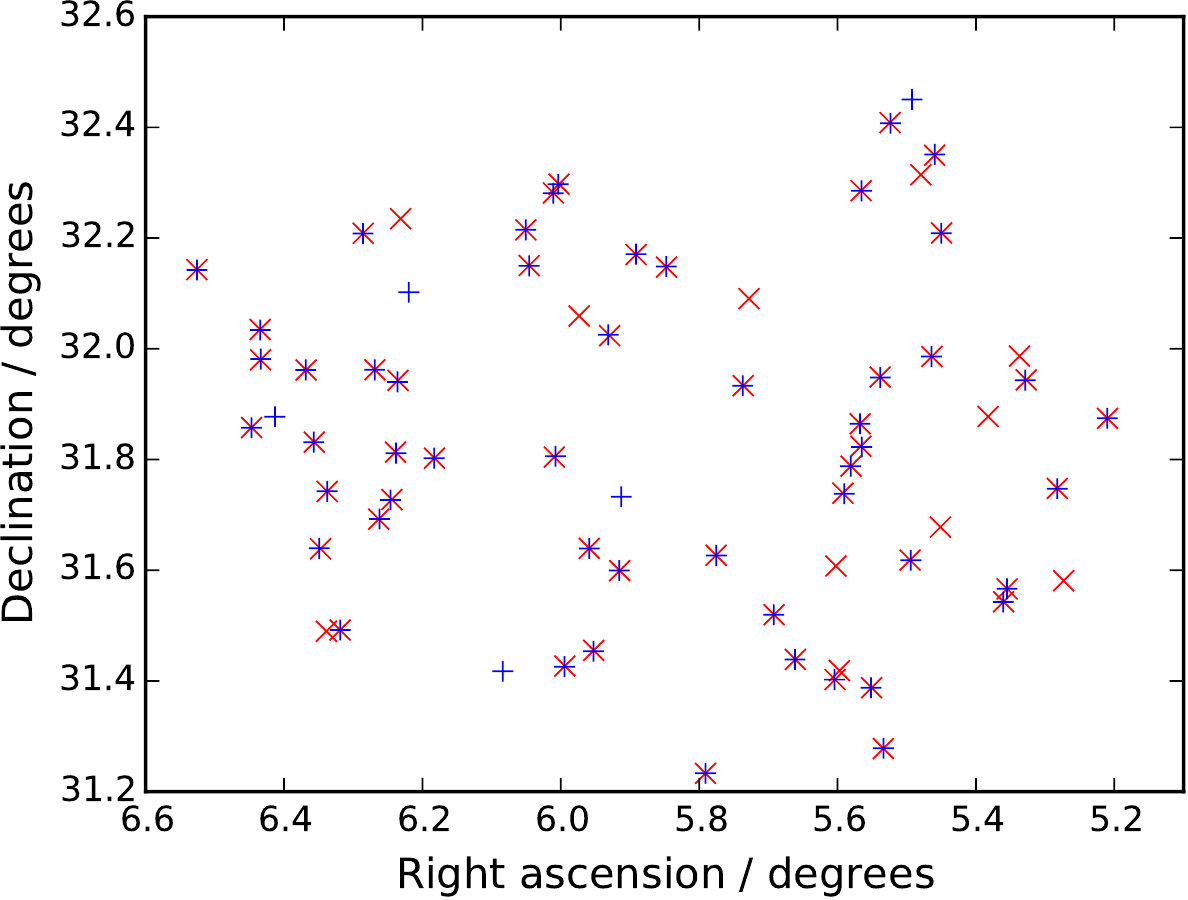}}
\caption{Positions of sources in the `first half' (red `$\times$') and `second half' (blue `+') catalogues in the AMI001 field when the two halves are imaged separately.}\label{fig:jackknife_positions}
\end{figure}

\begin{figure*}
\centerline{\includegraphics[width=\columnwidth]{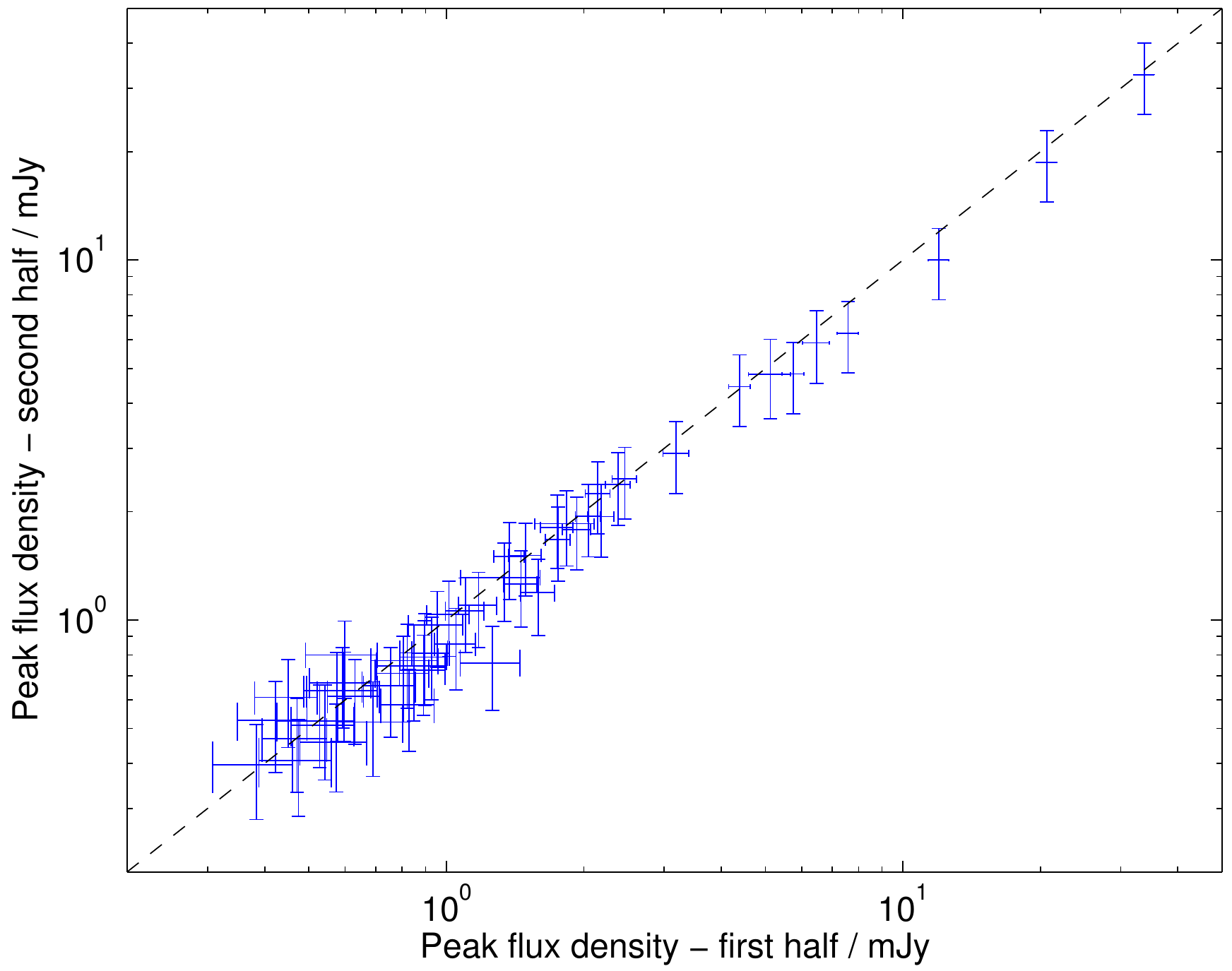}
			\quad
			\includegraphics[width=\columnwidth]{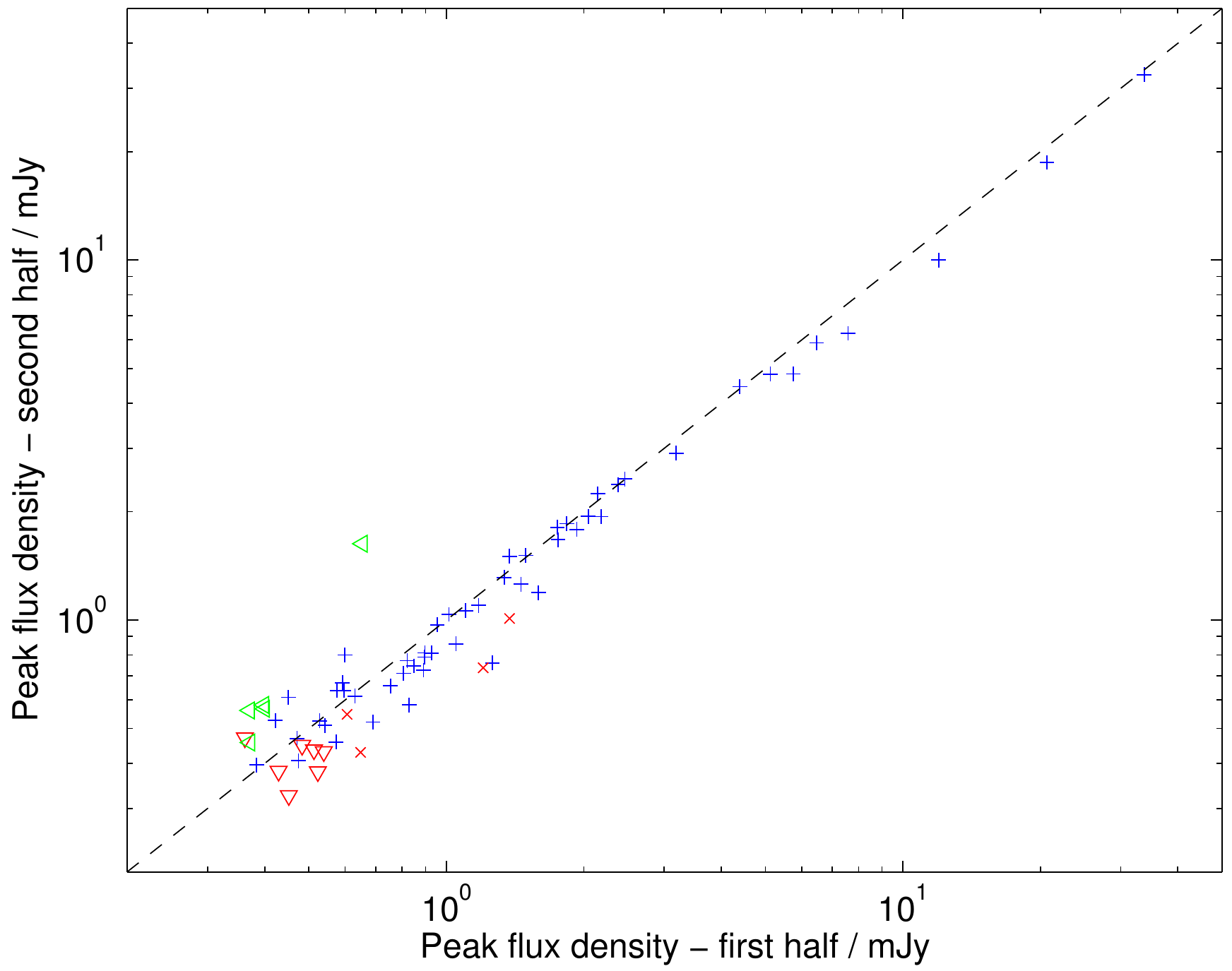}}
\caption{Peak flux densities of sources in the `first half' and `second half' catalogues in the AMI001 field, shown with error bars in the left panel, and without in the right panel for clarity. In the right panel upper limits are included for the 16 sources which are not found in one of the halves -- red `$\triangledown$' indicate upper limits for sources found in the first half and not in the second half, green `$\triangleleft$' indicate upper limits for sources found in the second half and not the first half. Red `$\times$' indicate the peak flux density of a source which was visible in the second half observations but which fell below the $5\sigma$ cut-off so was not included in the original catalogue.}\label{fig:jackknife_flux}
\end{figure*}

A long-term stability test was performed on a $\sim 1~{\rm deg}^2$ region in the deepest part of the AMI001 field. This involves splitting the data in half according to when it was observed, so all the older observations are in the first half and all the newer observations are in the second half. This was done in such a way as to ensure that the noise was the same in the two halves, so the split does not necessarily occur at the same point in time for every pointing. The median point in time for the split to occur is August 2009, but the exact date varies from pointing to pointing, because more observations were made at some pointings earlier or later during the observation period, or because some of the data were particularly noisy. The two halves are then imaged separately and \textsc{source\_find} was run on the two halves to create two catalogues. The `first half' catalogue contains 63 sources and the `second half' catalogue contains 57 sources; the positions of these sources are shown in Fig. \ref{fig:jackknife_positions}. The majority of sources are found in both catalogues, but 16 sources are only found in one or other of the catalogues. In order to compare the flux densities in the two halves, upper limits of five times the local noise are placed on the flux densities of these sources in the image they were not detected in; in four cases a source was visible in the image which had fallen below the $5\sigma$ cut-off and in these cases the peak flux density of the visible source was used instead. For five of the sources which are detected in both images there are small differences in the positions from the two halves; these sources all have low signal to noise values and the discrepancy is due to a different pixel being identified as the peak in the two halves. The flux densities of the sources in the two halves are compared in Fig. \ref{fig:jackknife_flux}. One source, which is not detected in the first half, is a significant outlier; this source is located very close to the edge of the map so could be spurious. There is some scatter in this plot but the majority of the values agree within the errors, so there do not appear to be any sources which are varying significantly on the timescales of several years in this sample. This suggests that there are few beamed objects amongst the flat-spectrum sources in this sample. There is also no evidence for any systematic differences between the data observed at the beginning and at the end of the observation run.

\section{Source counts}\label{section:source_counts}

\subsection{Calculating the source counts}

\begin{figure*}
\centerline{\includegraphics[width=\columnwidth]{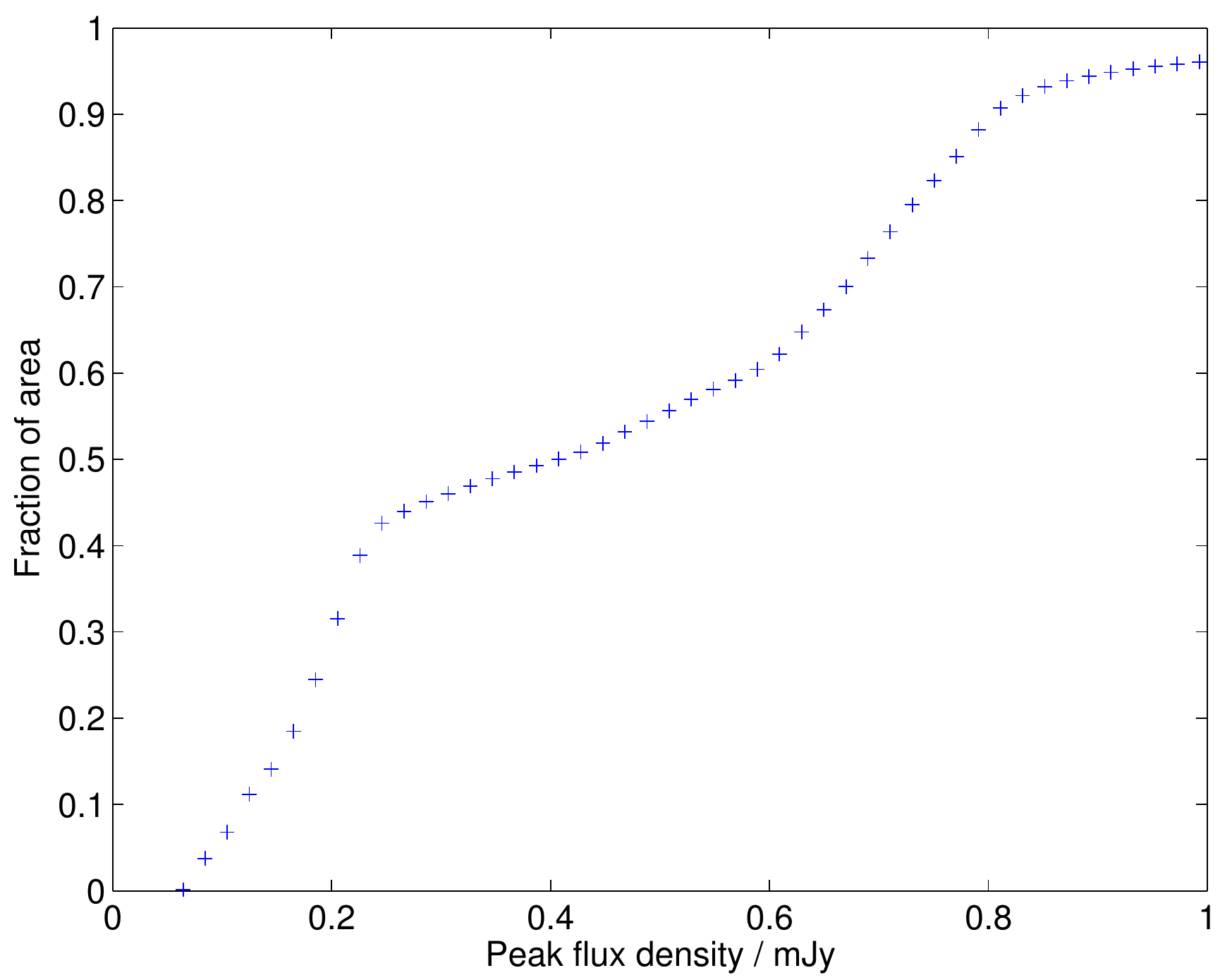}
			\quad
			\includegraphics[width=\columnwidth]{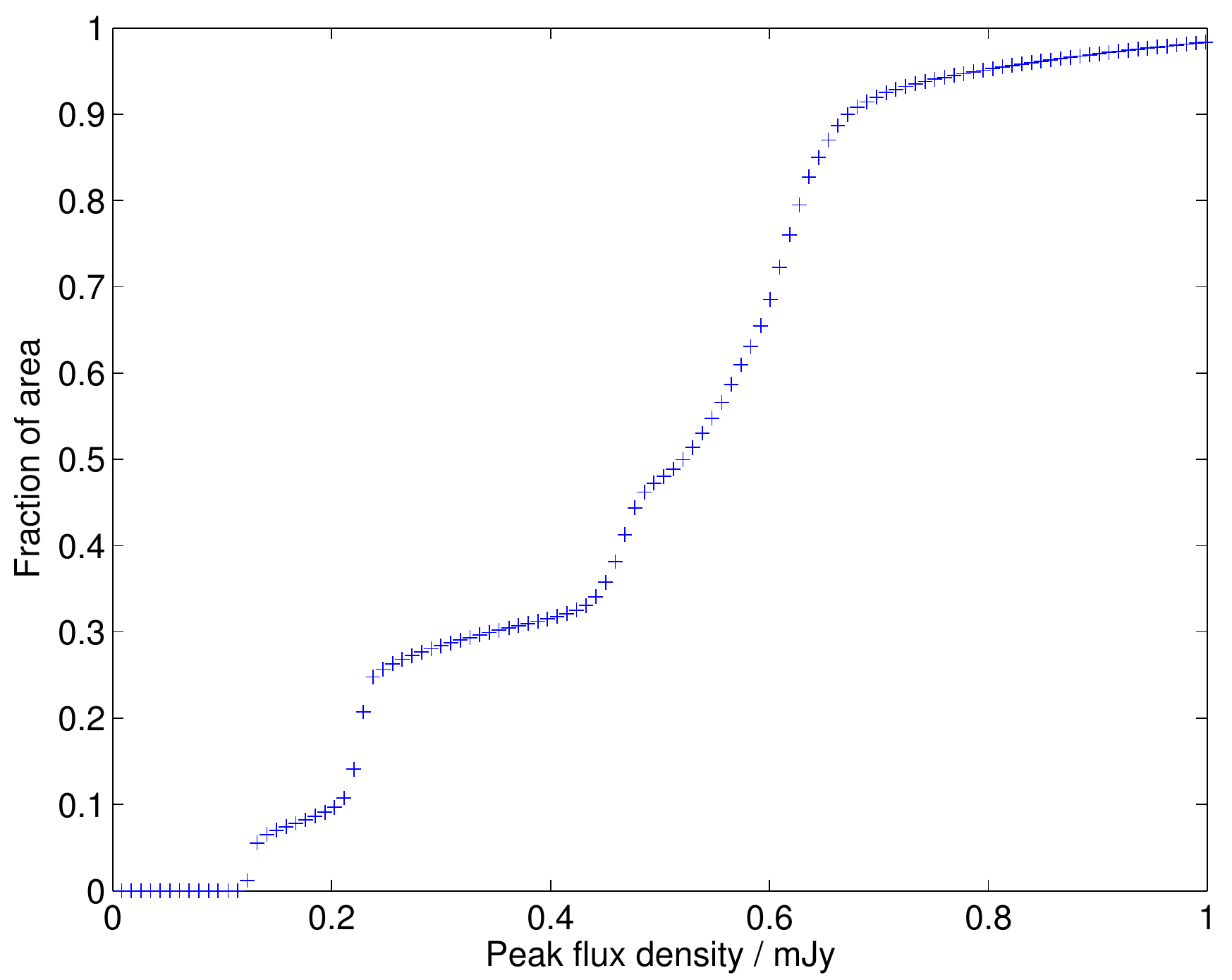}}
\caption{Visibility area of the AMI001 field (left) and Lockman Hole field (right) i.e.\ fraction of the total area over which a source with a given flux density could be detected.}\label{fig:vis_area}
\end{figure*}

The catalogue of sources used to calculate the source counts is described in Section~\ref{section:catalogue}. The integrated flux density is used for sources classified as extended in the fitting process, otherwise the peak flux density is used. The noise varies significantly across both fields, as is demonstrated in Figs.\ \ref{fig:AMI1_noise} and \ref{fig:LH_noise}, and this needs to be taken into account when calculating the source counts. To do this the visibility area was calculated \citep{1973A&A....23..171K}; this is the fraction of the total area over which a source of a flux density $S_i$ could be detected (i.e.\ the fraction of the total area with 5$\sigma_{\rm n} < S_i$), assuming the noise map. The visibility areas of the two fields are shown in Fig. \ref{fig:vis_area}. 

To calculate the source count, each source was weighted by the reciprocal of its visibility area. The source count in each flux density bin is therefore given by the following expression:
\begin{equation}
\frac{1}{A} \sum_{i=1}^{N} \frac{1}{x(S_i)}
\end{equation}
where $N$ is the number of sources in the bin, $A$ is the total area of the field and $x(S_i)$ is the visibility area for a source with flux density $S_i$. 

\begin{table}
\caption{Source counts in the AMI001 field.}\label{tab:AMI001-counts}
\centering
\medskip
\begin{tabular}{ccc}\\\hline
\vpad
Bin flux density  & ${\rm d}N/{\rm d}S$ \\
range / mJy       & / Jy$^{-1}$ sr$^{-1}$ \\\hline
 5.500  -- 9.000  & $ (9.3 \pm 4.7) \times 10^5 $ \\
 2.900  -- 5.500  & $ (6.7 \pm 1.5) \times 10^6 $ \\
 2.050  -- 2.900  & $ (2.0 \pm 0.4) \times 10^7 $ \\
 1.500  -- 2.050  & $ (3.6 \pm 0.8) \times 10^7 $ \\
 1.250  -- 1.500  & $ (6.7 \pm 1.5) \times 10^7 $ \\
 1.000  -- 1.250  & $ (8.3 \pm 1.7) \times 10^7 $ \\
 0.900  -- 1.000  & $ (1.3 \pm 0.3) \times 10^8 $ \\
 0.775  -- 0.900  & $ (1.2 \pm 0.3) \times 10^8 $ \\
 0.680  -- 0.775  & $ (1.8 \pm 0.5) \times 10^8 $ \\
 0.600  -- 0.680  & $ (2.3 \pm 0.6) \times 10^8 $ \\
 0.540  -- 0.600  & $ (2.7 \pm 0.8) \times 10^8 $ \\
 0.500  -- 0.540  & $ (4.3 \pm 1.3) \times 10^8 $ \\
 0.400  -- 0.500  & $ (4.2 \pm 0.8) \times 10^8 $ \\
 0.300  -- 0.400  & $ (5.9 \pm 1.0) \times 10^8 $ \\
 0.250  -- 0.300  & $ (9.1 \pm 1.9) \times 10^8 $ \\
 0.200  -- 0.250  & $ (1.7 \pm 0.3) \times 10^9 $ \\
 0.100  -- 0.200  & $ (2.5 \pm 0.4) \times 10^9 $ \\\hline
\end{tabular}
\end{table}

\begin{table}
\caption{Source counts in the Lockman Hole field.}\label{tab:LH-counts}
\centering
\medskip
\begin{tabular}{ccc}\\\hline
\vpad
Bin flux density  & ${\rm d}N/{\rm d}S$ \\
range / mJy       & / Jy$^{-1}$ sr$^{-1}$\\\hline
 2.900  -- 5.500    & $(3.7 \pm 1.5) \times 10^6$ \\
 2.050  -- 2.900    & $(2.3 \pm 0.6) \times 10^7$ \\
 1.500  -- 2.050    & $(2.4 \pm 0.8) \times 10^7$ \\
 1.250  -- 1.500    & $(3.4 \pm 1.5) \times 10^7$ \\
 1.000  -- 1.250    & $(9.7 \pm 2.6) \times 10^7$ \\
 0.900  -- 1.000    & $(7.1 \pm 3.5) \times 10^7$ \\
 0.775  -- 0.900    & $(1.9 \pm 0.5) \times 10^8$ \\
 0.680  -- 0.775    & $(1.8 \pm 0.6) \times 10^8$\\
 0.500  -- 0.680    & $(3.2 \pm 0.7) \times 10^8$\\
 0.300  -- 0.500    & $(4.3 \pm 1.1) \times 10^8$\\
 0.200  -- 0.300    & $(6.4 \pm 2.0) \times 10^8$\\
 0.120  -- 0.200    & $(1.9 \pm 0.7)\times 10^9$\\\hline
\end{tabular}
\end{table}

\begin{table}
\caption{Source counts for the two fields combined.}\label{tab:combined-counts}
\centering
\medskip
\begin{tabular}{ccc}\\\hline
\vpad
Bin flux density  & ${\rm d}N/{\rm d}S$ \\
range / mJy       & / Jy$^{-1}$ sr$^{-1}$ \\\hline
 2.900  -- 5.500  & $ (5.7 \pm 1.1) \times 10^6 $ \\
 2.050  -- 2.900  & $ (2.1 \pm 0.4) \times 10^7 $ \\
 1.500  -- 2.050  & $ (3.2 \pm 0.6) \times 10^7 $ \\
 1.250  -- 1.500  & $ (5.6 \pm 1.1) \times 10^7 $ \\
 1.000  -- 1.250  & $ (8.8 \pm 1.4) \times 10^7 $ \\
 0.900  -- 1.000  & $ (1.1 \pm 0.3) \times 10^8 $ \\
 0.775  -- 0.900  & $ (1.4 \pm 0.3) \times 10^8 $ \\
 0.680  -- 0.775  & $ (1.8 \pm 0.4) \times 10^8 $ \\
 0.600  -- 0.680  & $ (2.1 \pm 0.5) \times 10^8 $ \\
 0.540  -- 0.600  & $ (3.7 \pm 0.8) \times 10^8 $ \\
 0.500  -- 0.540  & $ (3.7 \pm 1.0) \times 10^8 $ \\
 0.400  -- 0.500  & $ (3.3 \pm 0.6) \times 10^8 $ \\
 0.300  -- 0.400  & $ (6.2 \pm 0.9) \times 10^8 $ \\
 0.250  -- 0.300  & $ (9.9 \pm 1.7) \times 10^8 $ \\
 0.200  -- 0.250  & $ (1.2 \pm 0.2) \times 10^9 $ \\
 0.100  -- 0.200  & $ (2.2 \pm 0.3) \times 10^9 $ \\\hline
\end{tabular}
\end{table}

\begin{figure}
\centerline{\includegraphics[width=\columnwidth]{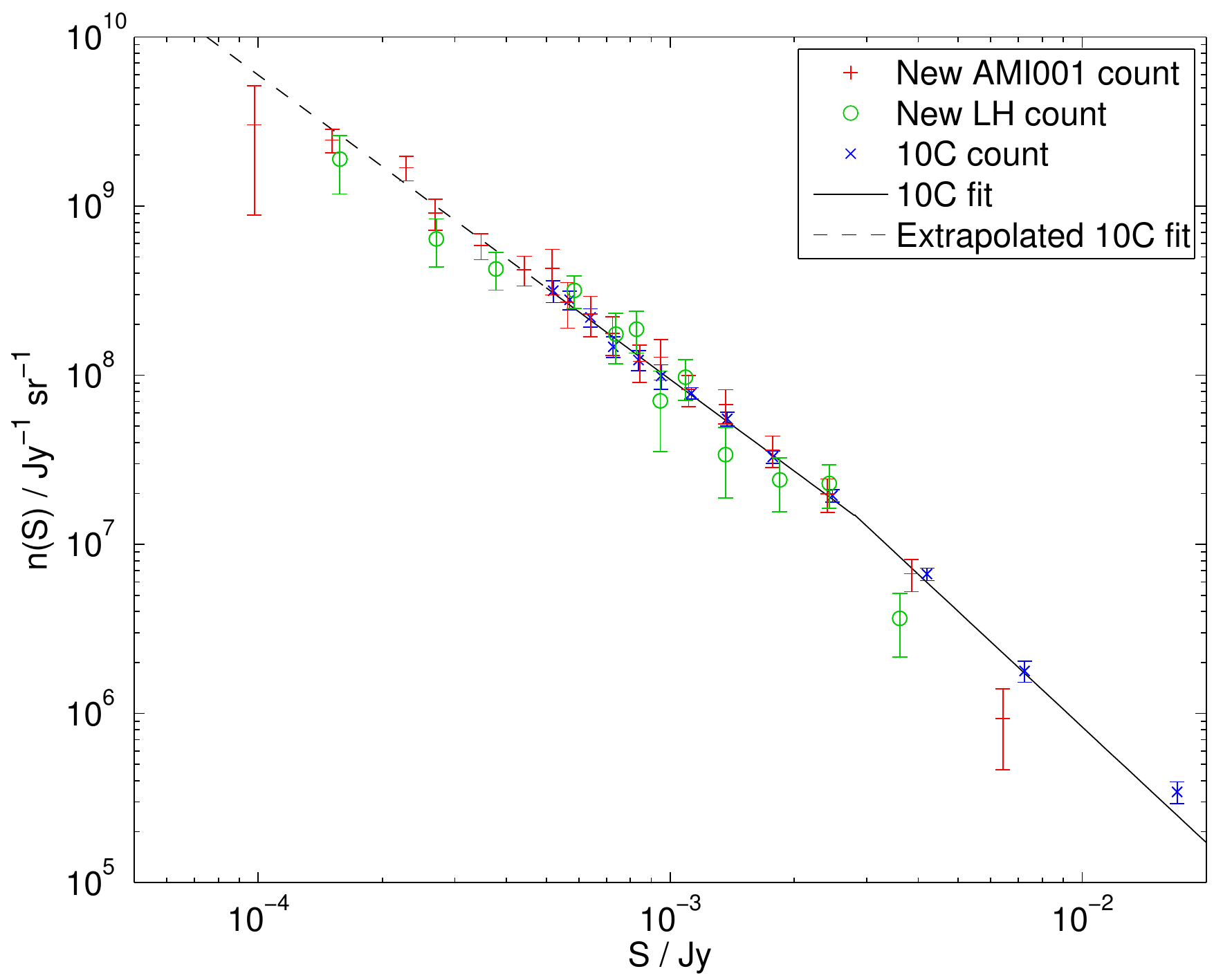}}
\caption{New source counts from the AMI001 (red `+') and Lockman Hole (green `$\circ$') fields. The 10C source counts are also shown (blue `$\times$') for comparison, as is the fit to the 10C source counts (black line). The faintest bin plotted for the AMI001 field count is based on only three sources and the completeness correction is not well defined at this flux density level, so this point is not included in the discussion or subsequent plots.}\label{fig:COUNTS}
\end{figure}

The differential source counts derived from the two fields are shown in Tables \ref{tab:AMI001-counts} and \ref{tab:LH-counts}, along with the combined count from the two fields in Table~\ref{tab:combined-counts}. They are plotted in Fig. \ref{fig:COUNTS}, along with the 10C count for comparison. The broken power-law fitted to the 10C count in Paper D, given by equation \ref{eqn:10C}, is also shown. The combined 9C, 10C and 10C ultra-deep (this work) counts are shown in Table~\ref{tab:9C-10C-10Cultra}.
\begin{equation}
%
n(S) \equiv \frac{{\rm d}N}{{\rm d}S} \approx \left\{
 \begin{array}{l l}
    24 \left(\frac{S}{\rm{Jy}}\right)^{-2.27} \textrm{Jy}^{-1} \textrm{sr}^{-1} &  \textrm{for 2.8 $\leq S \leq$ 25 mJy},\\
    376 \left(\frac{S}{\rm{Jy}}\right)^{-1.80} \textrm{Jy}^{-1} \textrm{sr}^{-1} & \textrm{for 0.5 $\leq S <$ 2.8 mJy}.\\
  \end{array} \right.
\label{eqn:10C}
\end{equation}  
Higher flux density bins are not included as the fields were chosen to lie away from bright sources. The points are plotted at the `centre of gravity' of each flux density bin (the average of the difference between each flux density and one edge of the bin). The error bars plotted are $\sqrt{N}$ Poisson errors. The new counts are in good agreement with the 10C count where they overlap, and extend the source count by a factor of five fainter in flux density. The counts from the two fields agree within the Poisson errors. The brightest flux density bins in the two fields appear to under-predict the 10C fit, this is probably due to the small number of sources in these bins, with four sources in the AMI001 field bin and six in the Lockman Hole field bin. The full 10C survey, which covers a much larger area, provides a more accurate measure of the source counts at flux densities greater than 0.5~mJy. In the faintest bin plotted for the AMI001 field ($0.08 < S_{15.7~\rm GHz} / \rm mJy < 0.10$) the completeness correction is not well defined, and due to the poor statistics in this bin it is omitted from Table \ref{tab:AMI001-counts} and later discussion.

\begin{table}
\caption{Source counts for the combined 9C, 10C and 10C ultra-deep counts.}\label{tab:9C-10C-10Cultra}
\centering
\medskip
\begin{tabular}{lcc}\\\hline
\vpad
Bin flux density  & ${\rm d}N/{\rm d}S$ \\
range / mJy       & / Jy$^{-1}$ sr$^{-1}$ \\\hline
500.0   -- 1000.0 & $ (1.01 \pm 0.36) \times 10^2 $ \\ 
200.0   -- 500.0  & $ (5.68 \pm 1.09) \times 10^2 $ \\ 
100.0   -- 200.0  & $ (2.97 \pm 0.43) \times 10^3 $ \\ 
60.0    -- 100.0  & $ (1.45 \pm 0.15) \times 10^4 $ \\ 
40.0    -- 60.0   & $ (3.06 \pm 0.31) \times 10^4 $ \\ 
30.0    -- 40.0   & $ (6.25 \pm 0.63) \times 10^4 $ \\ 
25.0    -- 30.0   & $ (1.00 \pm 0.11) \times 10^5 $ \\ 
16.0    -- 25.0   & $ (1.81 \pm 0.23) \times 10^5 $ \\ 
12.0    -- 16.0   & $ (4.22 \pm 0.53) \times 10^5 $ \\ 
10.0    -- 12.0   & $ (6.32 \pm 0.92) \times 10^5 $ \\ 
9.00    -- 10.0   & $ (1.03 \pm 0.27) \times 10^6 $ \\ 
6.40    -- 9.00   & $ (1.28 \pm 0.19) \times 10^6 $ \\ 
5.50    -- 6.40   & $ (3.40 \pm 0.51) \times 10^6 $ \\ 
2.90    -- 5.50   & $ (6.68 \pm 0.56) \times 10^6 $ \\ 
2.05    -- 2.90   & $ (1.94 \pm 0.17) \times 10^7 $ \\ 
1.50    -- 2.05   & $ (3.29 \pm 0.27) \times 10^7 $ \\ 
1.20    -- 1.50   & $ (5.70 \pm 0.48) \times 10^7 $ \\ 
1.00    -- 1.20   & $ (8.13 \pm 0.70) \times 10^7 $ \\ 
0.900   -- 1.00   & $ (9.88 \pm 1.65) \times 10^7 $ \\ 
0.775   -- 0.900  & $ (1.23 \pm 0.16) \times 10^8 $ \\ 
0.680   -- 0.775  & $ (1.47 \pm 0.20) \times 10^8 $ \\ 
0.600   -- 0.680  & $ (2.19 \pm 0.27) \times 10^8 $ \\ 
0.540   -- 0.600  & $ (2.79 \pm 0.36) \times 10^8 $ \\ 
0.500   -- 0.540  & $ (3.16 \pm 0.47) \times 10^8 $ \\ 
0.400   -- 0.500  & $ (3.30 \pm 0.61) \times 10^8 $ \\
0.300   -- 0.400  & $ (6.22 \pm 0.98) \times 10^8 $ \\
0.250   -- 0.300  & $ (9.89 \pm 1.75) \times 10^8 $ \\
0.200   -- 0.250  & $ (1.15 \pm 0.19) \times 10^9 $ \\
0.100   -- 0.200  & $ (2.23 \pm 0.33) \times 10^9 $ \\\hline
\end{tabular}
\end{table}

\subsection{Sample variance}

\begin{figure}
\centerline{\includegraphics[width=\columnwidth]{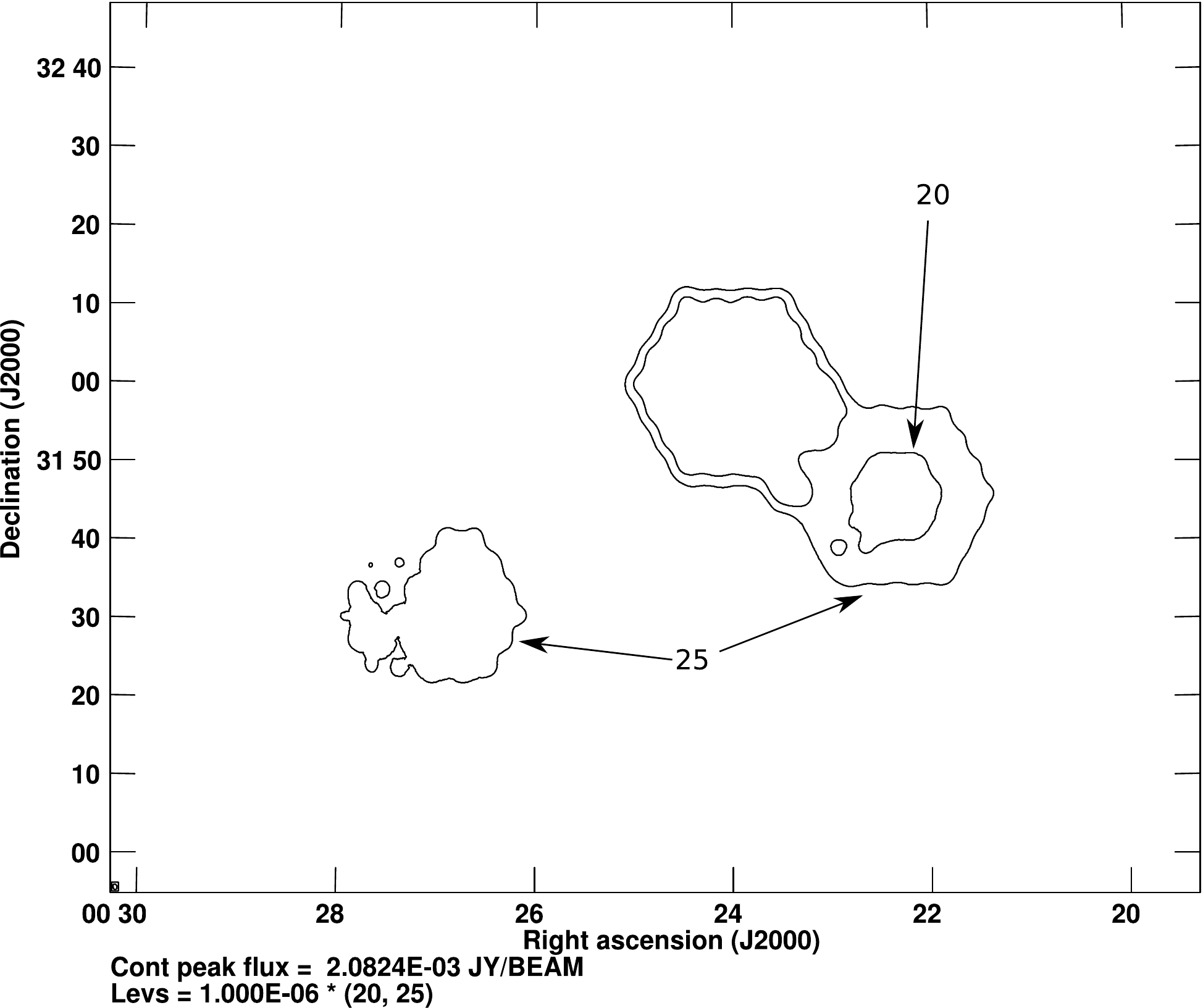}}
\caption{Contours at 20 and 25~$\muup$Jy, showning the regions in the AMI001 field with rms noise $< 25$ and 20~$\muup$Jy. }\label{fig:noise_kntrs_AMI001}
\end{figure}

\citet{2013MNRAS.432.2625H} investigated the effects on deep source counts at 1.4~GHz of sample variance induced by source clustering  by extracting a series of independent samples from the S$^3$ catalogue and comparing to observations. They used this to present a method for estimating the uncertainty in the source count caused by sample variance for an arbitrary radio survey. In the AMI001 field there is 0.23 deg$^2$ with rms noise $<20~\muup$Jy and 0.42 deg$^2$ with rms noise $<25~\muup$Jy, as shown in the contour plot in Fig. \ref{fig:noise_kntrs_AMI001}. It should therefore be possible to detect a source with $S_{15.7~\rm GHz} > 0.1$~mJy in an area of 0.23 deg$^2$; reading off Fig. 2 in \citet{2013MNRAS.432.2625H} for this survey limit and area gives an uncertainty due to cosmic variance of $\approx 14$~per cent. At higher flux densities, this uncertainty decreases as the area over which a source could be detected (the effective survey area) increases. For example, for a survey limit of 0.125~mJy (equivalent to an rms noise of 25$~\muup$Jy) and an area of 0.42 deg$^2$ the uncertainty is 12 per cent. We therefore expect cosmic variance to affect the faintest flux density bin by $\approx 14$ per cent.

\subsection{Possible biases}

Several effects which could bias the source counts are considered here; however for the reasons discussed no corrections are made for these biases.

\begin{enumerate}

\item Resolution bias. To calculate the source counts a sample which is complete in terms of integrated flux density is required, but the sources are detected according to their peak flux densities. This means that a resolved source of a given total flux density is more likely to fall below the peak flux density detection threshold than a point source with the same total flux density. Due to the relatively large beam size of AMI, only 7 per cent of sources in the AMI001 field and 4 per cent in the Lockman Hole field are extended, so resolution is not expected to have a significant effect on these source counts.

\item Eddington bias \citep{1913MNRAS..73..359E}. Statistical fluctuations due to thermal noise can alter the flux density of sources and can therefore cause some sources to be put into the wrong bins. Given the shape of the source count, if the true number of sources in one bin is larger than in adjacent bins, more sources will be scattered into that bin than out of it. The observed number of sources will therefore be biased high. In Paper D it is estimated that this effect will increase the number of sources in the faintest 10C flux density bin by $\approx 7$ per cent. As incompleteness is expected to reduce the number of sources in this bin by a similar amount they do not correct for this effect. No correction is applied here for the same reason.

\item Variability. Variability in flux densities can cause sources near the edge of flux density bins to move between bins. The shape of the source counts means that at the bottom of the bin, the number of sources in the positive phase of variability which are included in the bin will be marginally higher than the number of sources in the negative phase of variability which are excluded. The opposite effect occurs at the other end of the bin but will not be enough to offset this effect. Therefore, variability will boost the number of sources in a bin, causing a constant shift in the observed source count. The long timescale over which the observations were made means that the effects of short-term variability are generally averaged out, and as discussed in Section~\ref{section:checks}, we do not find any evidence for significant variability on longer timescales. 

\end{enumerate}

\section{Discussion}\label{section:sc_discussion}

\begin{figure*}
\centerline{\includegraphics[width=12cm]{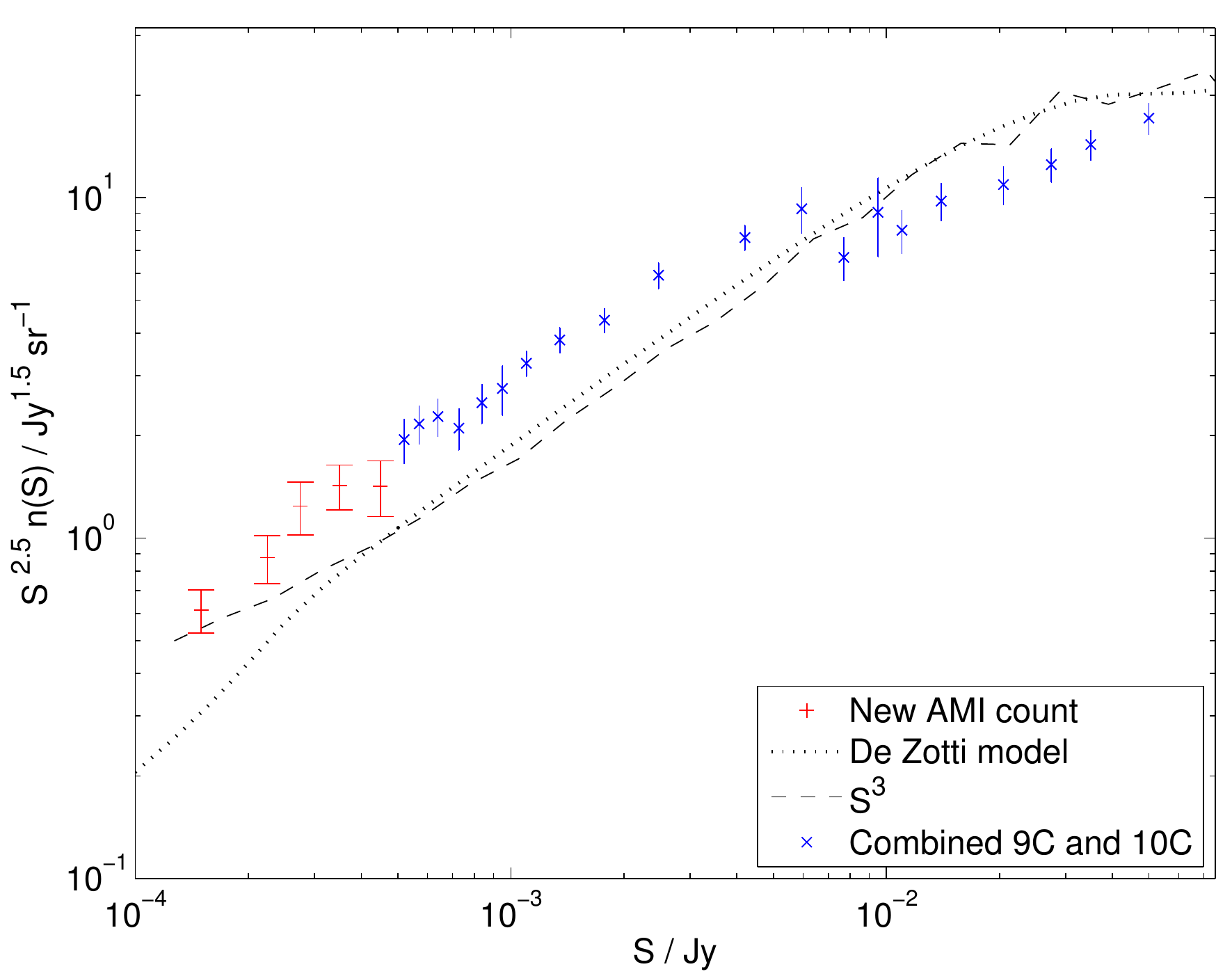}}
\caption{Euclidean normalised differential source counts from the new observations in both fields combined (red `+') and the combined 9C and 10C source counts from Paper D (blue `$\times$'). The \citeauthor{2005A&A...431..893D} model at 15~GHz (dotted line) and the 18~GHz count from the S$^3$ catalogue (dashed line) are also shown (no attempt is made to correct the latter to 15~GHz). Poisson errors are plotted for the observed counts.}\label{fig:de-zotti-counts-combined}
\end{figure*}

Figs. \ref{fig:COUNTS} and \ref{fig:de-zotti-counts-combined} show that the new deeper counts are consistent with the extrapolated 10C fit. There is no sign of the upturn observed in the 1.4-GHz source counts at $S_{1.4~\rm GHz} \sim 1$~mJy (e.g.\ \citealt{2010A&ARv..18....1D}); there is thus no evidence for a new population (e.g.\ of starforming galaxies) contributing to the 15.7-GHz source population above 0.1~mJy. This is not surprising, as a source with $\alpha = 0.8$, typical for a starforming galaxy \citep{2014MNRAS.439.1212F}, with $S_{1.4~\rm GHz} = 1$~mJy will have $S_{15.7~\rm GHz} \sim 0.14$~mJy, and would therefore only have appeared at the bottom of the faintest flux density bin here. 

A model of the high-frequency ($\nu > 5~\rm GHz$) source counts was produced by \citet{2005A&A...431..893D}, as described in Section~\ref{section:intro}. The new 15.7-GHz source counts presented here are compared to the latest version of the \citeauthor{2005A&A...431..893D} model, extracted from their website\footnote{http://web.oapd.inaf.it/rstools/srccnt\_tables}, in Fig. \ref{fig:de-zotti-counts-combined}. Paper D showed that the \citeauthor{2005A&A...431..893D} model under-predicts the number of sources in the 10C survey below $\approx 5$~mJy; it is clear in Fig. \ref{fig:de-zotti-counts-combined} that the model continues to under-predict the number of sources observed by a factor of two as flux density decreases. \citet{2013MNRAS.429.2080W} studied the spectral indices of a sample of 10C sources and showed that the proportion of flat-spectrum sources in particular is too low in the \citeauthor{2005A&A...431..893D} model below $\approx 1$~mJy; the model predicts that at $S_{15~\rm GHz} = 1$~mJy steep-spectrum sources outnumber flat-spectrum source by a factor of three, while the observations show that there are twice as many flat-spectrum sources as steep-spectrum sources. It is likely that this under-prediction of the number of flat-spectrum sources in the sub-mJy population is responsible for the discrepancies between the model and the observed count seen here at $S < 5~\rm mJy$. Above 5~mJy the model over-predicts the number of sources observed, this is discussed further in Paper D. The higher-flux density end (0.5 mJy to several Jy) of the 15-GHz source counts is studied by combining the 9C and 10C counts with the AT20G survey by \citet{2014MNRAS.439.1212F}, who investigated the source counts of the steep and flat-spectrum populations separately. These counts are compared to the \citeauthor{2005A&A...431..893D} and \citet{1999MNRAS.304..160J} models, and they find that both underestimate the number of flat-spectrum sources below 5~mJy.

The new 15.7-GHz source count is also compared to the S$^3$ catalogue in Fig. \ref{fig:de-zotti-counts-combined}. All sources with $S_{18~\rm GHz} > 0.09$~mJy were selected from the simulation, and the source count was calculated in the same bins as for the observed count. The simulation under-predicts the observed number of sources in a similar way to the \citeauthor{2005A&A...431..893D} model. It is likely that this under-prediction is again due to a lack of flat spectrum sources in the model, as \citeauthor{2013MNRAS.429.2080W} showed that the S$^3$ and 10C spectral index distributions are significantly different, with the simulation missing almost all the flat spectrum sources. (Note that the S$^3$ and \citeauthor{2005A&A...431..893D} models are not entirely independent as they are both extrapolations from models constructed using low-frequency data.) \citet{whittaminpress} used multi-wavelength data to show that these flat-spectrum 10C sources are probably the result of emission from the cores of radio galaxies, which therefore have a far greater contribution than predicted by either of the models discussed here.

Below $\approx 0.3$~mJy there is a better agreement between the observed and simulated counts due to a slight flattening in the S$^3$ counts. This flattening is due to the greater contribution of starforming sources (both quiescent and starbursting) to the simulated catalogue below $\approx 0.3$~mJy; starforming sources comprise 21 per cent of the simulated sources with $0.09 < S_{18~\rm GHz} / \rm mJy < 0.3$, compared to only 7 per cent of sources with $S_{18~\rm GHz} > 0.5$~mJy. However, given that there is no flattening in the new AMI count, it is not clear what contribution, if any, a population of starforming sources is making to the observed counts at $S_{15.7~\rm GHz} > 0.1$~mJy. 

A study of the multi-wavelength properties of 10C sources in the Lockman Hole \citep{whittaminpress} used radio-to-optical ratios to show that at least 94 per cent of the 10C sample ($S_{15~\rm GHz} > 0.5$~mJy) are radio loud, demonstrating that there is no significant population of starforming galaxies present at these flux density levels. Given that there is no change in the slope of the source count for $0.1 < S/ \rm mJy < 0.5$, it seems unlikely that starforming galaxies are making a significantly greater contribution in this flux density range than at higher flux densities. Therefore, the population of starforming galaxies predicted to be making a significant contribution ($\sim 20$ per cent) in this flux density range by the S$^3$ simulation do not appear to be present. This is consistent with the results of several recent studies of the faint ($S_{1.4~\rm GHz} < 0.1$~mJy) source population at lower frequencies \citep{2012MNRAS.421.3060S,2014MNRAS.440.1527L,2015arXiv150701144L} which have also found fewer starforming galaxies than predicted by the simulation.

\section{Conclusions}\label{section:conclusions}

This paper presents new very deep (best rms noise = 16~$\muup$Jy beam$^{-1}$) 15.7-GHz observations in two fields. These are the deepest high-frequency radio observations to date, and enable us to calculate the source counts down to $S_{15.7~\rm GHz} = 0.1$~mJy. This is a factor of five deeper than previously achieved with the 10C survey.

The source counts are consistent with the extrapolated fit to the 10C count. There is thus no evidence for a new population of objects contributing to the 15.7-GHz source counts above 0.1~mJy, suggesting that the high-frequency radio sky continues to be dominated by radio galaxies down to at least this flux density level. We do not observe the population of starforming galaxies predicted to begin to contribute $0.1 < S/ \rm mJy < 0.5$ by the S$^3$ simulation.

Comparisons with the \citeauthor{2005A&A...431..893D} model and S$^3$ simulation show that both these underestimate the observed number of sources at low flux densities by a factor of two. This is probably due to the flat-spectrum cores of radio galaxies contributing more significantly to the counts than predicted by the models.

\section*{Acknowledgements}

We thank the staff of the Mullard Radio Astronomy Observatory for maintaining and operating AMI. IHW and CR acknowledge Science and Technology
Facilities Council studentships. IHW acknowledges support from the Square Kilometre Array South Africa project and the South African National Research Foundation. This research has made use of NASA's Astrophysics Data System. We thank the referee for their careful reading of this manuscript. 

%
%

\setlength{\labelwidth}{0pt} 

\bsp

\label{lastpage}

\begin{thebibliography}{}

 \bibitem[AMI Consortium: Davies et al.(2011)Davies et al.]{2011MNRAS.415.2708D} 
  AMI Consortium: Davies, et al., 2011, MNRAS, 415, 2708 (Paper D)

 \bibitem[AMI Consortium: Franzen et al.(2011)Franzen et al.]{2011MNRAS.415.2699F} 
  AMI Consortium: Franzen, et al., 2011, MNRAS, 415, 2699 (Paper F)

 \bibitem[de Zotti et al.(2005)de Zotti et al.]{2005A&A...431..893D} 
  de Zotti G., Ricci R., Mesa D., Silva L., Mazzotta P., Toffolatti L., Gonz{\'a}lez-Nuevo J., 2005, A\&A, 431, 893 

 \bibitem[de Zotti et al.(2010)de Zotti et al.]{2010A&ARv..18....1D} 
  de Zotti G., Massardi M., Negrello M., Wall J., 2010, A\&ARv, 18, 1 

 \bibitem[Dunlop \& Peacock(1990)Dunlop \& Peacock]{1990MNRAS.247...19D} 
  Dunlop J.~S., Peacock J.~A., 1990, MNRAS, 247, 19 

 \bibitem[Eddington(1913)Eddington]{1913MNRAS..73..359E} 
  Eddington A.~S., 1913, MNRAS, 73, 359 

 \bibitem[Fanaroff \& Riley(1974)Fanaroff \& Riley]{1974MNRAS.167P..31F} 
  Fanaroff B.~L., Riley J.~M., 1974, MNRAS, 167, 31P 

 \bibitem[Franzen et al.(2014)Franzen et al.]{2014MNRAS.439.1212F} 
  Franzen T.~M.~O., et al., 2014, MNRAS, 439, 1212 

 \bibitem[Heywood et al.(2013)Heywood, Jarvis, \& Condon]{2013MNRAS.432.2625H} 
  Heywood I., Jarvis M.~J., Condon J.~J., 2013, MNRAS, 432, 2625 

 \bibitem[Jackson \& Wall(1999)Jackson \& Wall]{1999MNRAS.304..160J} 
  Jackson C.~A., Wall J.~V., 1999, MNRAS, 304, 160 

 \bibitem[Katgert et al.(1973)Katgert et al.]{1973A&A....23..171K} 
  Katgert P., Katgert-Merkelijn J.~K., Le Poole R.~S., van der Laan H., 1973, A\&A, 23, 171 

 \bibitem[Lindsay et al.(2014)Lindsay et al.]{2014MNRAS.440.1527L} 
  Lindsay S.~N., et al., 2014, MNRAS, 440, 1527 

 \bibitem[Luchsinger et al.(2015)Luchsinger et al.]{2015arXiv150701144L} 
  Luchsinger K.~M., et al., 2015, arXiv, arXiv:1507.01144 

 \bibitem[Massardi et al.(2011)Massardi et al.]{2011MNRAS.412..318M} 
  Massardi M., et al., 2011, MNRAS, 412, 318 

 \bibitem[Simpson et al.(2012)Simpson et al.]{2012MNRAS.421.3060S} 
  Simpson C., et al., 2012, MNRAS, 421, 3060 

 \bibitem[Toffolatti et al.(1998)Toffolatti et al.]{1998MNRAS.297..117T} 
  Toffolatti L., Argueso Gomez F., de Zotti G., Mazzei P., Franceschini A., Danese L., Burigana C., 1998, MNRAS, 297, 117 

 \bibitem[Tucci et al.(2011)Tucci et al.]{2011A&A...533A..57T} 
  Tucci M., Toffolatti L., de Zotti G., Mart{\'{\i}}nez-Gonz{\'a}lez E., 2011, A\&A, 533, A57 

 \bibitem[Waldram et al.(2003)Waldram et al.]{2003MNRAS.342..915W} 
  Waldram E.~M., Pooley G.~G., Grainge K.~J.~B., Jones M.~E., Saunders R.~D.~E., Scott P.~F., Taylor A.~C., 2003, MNRAS, 342, 915 

 \bibitem[Waldram et al.(2010)Waldram et al.]{2010MNRAS.404.1005W} 
  Waldram E.~M., Pooley G.~G., Davies M.~L., Grainge K.~J.~B., Scott P.~F., 2010, MNRAS, 404, 1005 

 \bibitem[Whittam et al.(2013)Whittam et al.]{2013MNRAS.429.2080W} 
  Whittam I.~H., et al., 2013, MNRAS, 429, 2080 

 \bibitem[Whittam et al.(2015)Whittam et al.]{whittaminpress} 
  Whittam I.~H., Riley J.~M., Green D.~A., Jarvis M.~J., Vaccari, M., 2015, MNRAS, in press

 \bibitem[Wilman et al.(2008)Wilman et al.]{2008MNRAS.388.1335W} 
  Wilman R.~J., et al., 2008, MNRAS, 388, 1335 

 \bibitem[Wilman et al.(2010)Wilman et al.]{2010MNRAS.405..447W} 
  Wilman R.~J., Jarvis M.~J., Mauch T., Rawlings S., Hickey S., 2010, MNRAS, 405, 447 

 \bibitem[Zwart et al.(2008)Zwart et al.]{2008MNRAS.391.1545Z} 
  Zwart J.~T.~L., et al., 2008, MNRAS, 391, 1545 


\end{thebibliography}
\end{document}